\documentclass[twocolumn,superscriptaddress,10pt]{revtex4-1}
\usepackage[T1]{fontenc}
\usepackage{times}
\usepackage[utf8]{inputenc}
\usepackage[resetlabels,labeled]{multibib}
\usepackage{amsmath}
\usepackage{amsthm}
\usepackage{amsfonts}
\usepackage{amssymb}
\usepackage{bbm}
\usepackage{bbold}
\usepackage{graphicx,epsfig,color,tcolorbox}
\usepackage{physics}
\usepackage{mathtools}
\usepackage{mathrsfs}
\usepackage[colorlinks=true,linkcolor=teal,citecolor=purple]{hyperref}
\usepackage{hyperref}
\usepackage{cleveref}

\newcommand{\Lam}{\mathbb{\Lambda}}

\newcommand{\ot}{\otimes}

\newcommand{\cptp}{{\large\texttt{CPTP}}}
\newcommand{\sep}{{\large\texttt{SEP}}}
\newcommand{\unitary}{{\large\mathsf{U}}}

\newcommand{\povm}{{\large\texttt{POVM}}}

\begin{document}

\title{The operational advantages provided by non-classical teleportation}

\author{Patryk Lipka-Bartosik}
 \affiliation{H. H. Wills Physics Laboratory, University of Bristol, Tyndall Avenue, Bristol, BS8 1TL, United Kingdom}
\author{Paul Skrzypczyk}
\affiliation{H. H. Wills Physics Laboratory, University of Bristol, Tyndall Avenue, Bristol, BS8 1TL, United Kingdom}

\date{\today}

\begin{abstract}

The standard benchmark for teleportation is the average fidelity of teleportation and according to this benchmark not all states are useful for teleportation. It was recently shown however that all entangled states lead to non-classical teleportation, with there being no classical scheme able to reproduce the states teleported to Bob. Here we study the operational significance of this result. On the one hand we demonstrate that every entangled state is useful for teleportation if a generalisation of the average fidelity of teleportation is considered which concerns teleporting quantum correlations. On the other hand, we show the strength of a particular entangled state and entangled measurement for teleportation -- as quantified by the robustness of teleportation -- precisely characterises their ability to offer an advantage in the task of subchannel discrimination with side information. This connection allows us to prove that every entangled state outperforms all separable states when acting as a quantum memory in this discrimination task. Finally, within the context of a resource theory of teleportation, we show that the two operational tasks considered provide complete sets of monotones for two partial orders based upon the notion of teleportation simulation, one classical, and one quantum.

\end{abstract}

\keywords{}
\maketitle

\section{Introduction}
Quantum teleportation \cite{Bennet1993} 
is one of the most important protocols in quantum information theory. In its standard form it involves transferring an unknown quantum state to a remote recipient using classical communication and pre-shared entanglement. 
Although nothing actually moves during the process, the situation can't be meaningfully distinguished from one in which the original state has been transported to another location. 
To date it has been demonstrated in a wide range of experiments \cite{Bouwmeester1997,Boschi1998,Furusawa1998,Bao2012,Leuenberger2005,Pirandola2015,Vaidman1994,Sherson2006} and is currently one of the building blocks in many quantum information contexts, ranging from distributed quantum networks \cite{Briegel1998}, quantum repeaters \cite{Hasegawa2019}, quantum computers \cite{Gottesman1999} and even the future quantum internet \cite{Kimble2008}.

In the ideal version of teleportation Alice and Bob share a maximally entangled state and Alice is given a system in some unknown state. She performs a Bell-state measurement on the system and her share of the entangled state and communicates the result to Bob who applies an appropriate unitary correction to his share and transforms it into the state given to Alice. 

However, in realistic teleportation protocols the states and measurements used are never perfect. This motivates studying a more general teleportation scheme involving arbitrary states and measurements. We will adapt this approach here and assume that Alice and Bob share an arbitrary quantum state and introduce a third party, called the Verifier, who gives Alice states to be teleported. She then applies an arbitrary measurement on her share of the entangled state and the system given to her and communicates the measurement result to Bob, who performs a local correction on his state.

The standard figure of merit used to quantify how well a given teleportation protocol performs is the \emph{average fidelity of teleportation}, denoted here by $\langle F \rangle$ and defined as the fidelity between the state to be teleported and the final state of Bob's after the protocol is finished, averaged uniformly over all measurement results and input states. This quantity was first introduced in \cite{Popescu1994} and since then has been used widely to quantify the usefulness of states for teleportation \cite{Horodecki1999,Linden1999,Olmschenk2009}. The average fidelity of teleportation is maximal when teleportation is perfect, i.e.~as in the ideal version. If Alice and Bob do not share an entangled state, or are unable to perform an entangled measurement, then the corresponding teleportation scheme is said to be ``classical''. For all such schemes the average fidelity can never exceed the threshold value $\langle F_{\text{c}} \rangle = 2/(d+1)$ \cite{Horodecki1999}, where $d$ is the local dimension of the shared state. Importantly, it was shown that there exist entangled states in Nature (e.g. bound-entangled states \cite{Linden1999,Bennet1999,HorodeckiP2003,Horodecki1998}) which cannot surpass this classical threshold. This led to a common belief that not all entangled states are useful for quantum teleportation. 

However, it was recently shown that the average fidelity is not sufficiently sensitive to probe all aspects of teleportation experiments \cite{Cavalcanti2017,Supic2019}. In particular, every entangled state can lead to non-classical teleportation if the full data from the experiment is taken into account \cite{Cavalcanti2017}. To show this a geometric method of quantifying the non-classicality of teleportation data using a measure called the \emph{robustness of teleportation} (RoT) was introduced. By showing that the RoT is non-zero whenever Alice and Bob share entanglement and Alice performs a Bell state measurement, it was demonstrated that every entangled state leads to experimental data which could not be produced without entanglement. However, the question of in what sense this non-classical data showed that the entanglement could be considered as being ``useful'' for teleportation in some operational sense has remained unanswered. 


In this work we construct a resource theory of quantum teleportation. Unlike other resource-theoretic studies in literature which address a single type of resource, quantum teleportation combines two distinct resources - shared entanglement and entangled measurement. Using this framework we show that RoT admits two natural operational interpretations. 

Firstly, it quantifies the advantage enabled by an entangled state and entangled measurement in the task of teleporting unknown quantum correlations -- rather than unknown states -- over all classical instruments. This task can be thought of as a natural generalization of entanglement swapping \cite{Zukowski1993,Pan1998} where the goal is not only to ''swap`` entanglement but to achieve pre-defined quantum correlations between parties. We show that the average score in this task when teleporting classical correlations reduces to the average fidelity of teleportation. This also shows a surprising property of bound-entangled states \cite{Horodecki1998} (i.e. states from which no entanglement can be distilled) -- they provide advantage over separable states in teleporting genuine quantum correlations. This also answers an open problem from \cite{Cavalcanti2017} by specifying in what sense all entangled states are useful for teleportation.

Secondly, we show that RoT also quantifies the maximal achievable advantage in the task of subchannel discrimination with quantum side information. This reveals that RoT is another robustness-based quantifier which fits into the program of discrimination tasks, a class of problems with fundamental importance to the field of quantum information \cite{Kitaev1997,Acin20011,Childs2000}. Analogous results have been shown also for entanglement \cite{Vidal1999,Bae2019,Takagi2019}, coherence \cite{Napoli2016}, EPR-steering \cite{Piani2015}, quantum measurement \cite{Skrzypczyk2019,Ducuara2019,Oszmaniec2019operational}, measurement incompatibility \cite{Designole2018,Designolle_2019} and fault-tolerant quantum computation \cite{Howard2017}. This surprising connection allows us to infer that every entangled state can act as a useful quantum memory for local subchannel discrimination.

Finally, by formulating teleportation in the language of resource theories, we show that both tasks provide complete sets of monotones for two natural notions of simulation (free operations), one classical and the other quantum.

\section{Framework}
We denote the set of all quantum channels by \cptp \, and the identity map with $\mathcal{I}$. An instrument $\mathbb{E} = \{\mathcal{E}_a\}$ is a collection of completely positive and trace non-increasing linear maps $\mathcal{E}_a$, so-called subchannels, such that $\sum_a \mathcal{E}_a[\cdot]$ is a channel. This naturally captures the concept of branching of a linear evolution \cite{Davies1970,Bae2019} and allows one to calculate both the (potentially state-dependent) probability $p(a) = \tr \mathcal{E}_a[\rho]$ of different branches acting on state $\rho$ and the corresponding final state $\mathcal{E}_a[\rho] / \tr \mathcal{E}_a[\rho]$.  

In our study of teleportation we will assume that Alice and Bob share an arbitrary quantum state $\rho^{\text{AB}}$ of dimension $d_{\text{A}} \times d_{\text{B}}$ and the third party, called the Verifier, provides quantum states $\{\omega_x^{\text{V}}\}$, $x = 0, 1, \ldots, n$ of dimension $d_{\text{V}}$, unknown to Alice. She then applies a general POVM (Positive Operator-Valued Measure) measurement $M_{a}^{\text{VA}}$ on her share of the entangled state and input system, as a result projecting Bob's state into:      
\begin{align}
    \rho_{a|\omega_x}^{\text{B}} = \frac{1}{p(a|x)} \tr_{\text{VA}} \left[(M_a^{\text{VA}} \ot \mathbb{1}^{\text{B}}) \left(\omega_x^{\text{V}} \ot \rho^{\text{AB}} \right)\right],
\end{align}
where $p(a|x) = \tr\left[(M_a^{\text{VA}} \ot \mathbb{1}^{\text{B}}) \left(\omega_x^{\text{V}} \ot \rho^{\text{AB}} \right)\right]$ is the probability of a particular outcome $a$ given that state $\omega_x$ was provided by the Verifier. For our purposes it will be more convenient to work with unnormalized states and thus we define:
\begin{align}
\label{eq:telep_instr}
%
\Lambda_a[\omega_x] := p(a|x) \cdot \rho_{a|\omega_x}^{\text{B}} = \sigma_{a|\omega_x}^{\text{B}},
\end{align}
where each $\Lambda_a\left[\cdot\right] = \Lambda_a^{\text{V}\rightarrow \text{B}}\left[\cdot\right]$ is a subchannel from V to B which transforms the input states $\omega_x$ into (unnormalised) output states $\sigma_{a|\omega_x}$. We will refer to such a collection as a teleportation instrument and denote it with $\Lam{} = \{\Lambda_a \}$. Notice that $\{M_a^{\text{VA}}\}$ form a POVM and hence $\Lam{}$ satisfies:
\begin{align}
\label{eq:14}
\sum_a \Lambda_a (\omega) = \rho^{\text{B}},
\end{align} 
irrespective of $\omega$, which can be interpreted as a no-signaling condition.

When the states $\{\omega_x\}$ form a tomographically-complete set, the experiment becomes effectively independent of the input (see the Appendix). This means that full information about teleportation instrument can be obtained by probing it with $\{\omega_x\}$ and motivates introducing a notion of a \emph{complete teleportation experiment}, i.e. an experiment in which the set of input states is tomographically-complete. In the remainder of this paper, we will focus exclusively on complete teleportation experiments.

Consider now the case when $\rho^{\text{AB}}$ is a separable state, i.e.  $\rho^{\text{AB}} = \sum_{\lambda} p_{\lambda} \, \rho_{\lambda}^{A} \ot \rho_{\lambda}^{\text{B}}$ and denoted by $\rho^{\text{AB}} \in \sep$. The associated teleportation instrument takes the form: 
\begin{align}
\Lambda^c_a (\omega_x) &= \nonumber \sum_{\lambda} p_{\lambda} \tr_{\text{VA}} \left[ \left(M_a^{\text{VA}} \ot \mathbb{1}^{\text{B}}\right) \left(\omega_x \ot \rho_{\lambda}^{\text{A}} \ot \rho_{\lambda}^{\text{B}} \right) \right] \\ \label{eq:16}
&= \sum_{\lambda} p_{\lambda}\, p(a|x, \lambda) \, \rho_{\lambda}^{\text{B}},
\end{align}
where $p(a|x, \lambda) = \tr[M_a^{\text{VA}} (\omega_x^{\text{V}} \ot \rho_{\lambda}^{\text{A}})]$. This is the most general classical teleportation scheme which can be realized if Alice and Bob have access only to classical randomness $\lambda$ and the ability to locally prepare quantum states in their labs. We will denote the set of all such teleportation instruments by $\mathcal{F}$, in analogy with the set of free objects studied in the context of resource theories \cite{Brandao2015,Bennett1996,Gour2008,Marvian2013,Aberg2006,Baumgratz2014,Horodecki2003a,Janzing2000,Horodecki2013,Brandao2013,Veitch2014,deVicente2014,Gallego2015,Horodecki2015,Chitambar2019,Liu2019,Liu2019op,Theurer2019}. If the teleportation data $\{\sigma_{a|\omega_x}^\mathrm{B}\}$ cannot be explained as coming from a classical teleportation instrument, we will refer to the associated teleportation instrument as ``quantum'' and denote the set of all such instruments with $\mathcal{R}$. 

In the standard approach the quality of a given teleportation instrument is assessed using the average fidelity of teleportation \cite{Popescu1994}, which in the present context is given by:
\begin{align}
    \label{eq:65}
    \langle F \rangle = \max_{\{U_a\}_a \in \unitary} \,\, \frac{1}{n}\sum_{a,x} p(a|x) \langle \omega_x | U_a \rho_{a|\omega_x}^{\text{B}} U_a^{\dagger}|\omega_x \rangle, 
\end{align}
where the maximisation is over all correcting unitaries $\{U_a\}$ for Bob, denoted $\unitary$. This quantity does not utilize all the data produced in the teleportation experiment. A method for quantifying how `close' a set of data is to that which could arise from a classical teleportation instrument is to solve the following convex optimization problem:
\begin{align} 
\label{eq:5}
\mathcal{T}(\mathbb{\Lambda}) :=  \min_{\{\Lambda_a^c\}, \{\Lambda_a'\}, r} \,\, & r, \\ \nonumber
    \text{s.t.} \quad \quad &  \frac{1}{1+r}\, \Lambda_{a}[\omega_x] + \frac{r}{1+r} \Lambda'_a[\omega_x] = \Lambda^c_a[\omega_x], \\ \nonumber
    & \{\Lambda^c_a\} \in \mathcal{F},\quad \{\Lambda_a'\} \in \mathcal{R}.
\end{align}
where $\Lambda'_a[\omega_x]$ describes the ``noise'' which comes from some other teleportation instrument $\Lam{}'$ and which has to be added to the teleportation data $\sigma_{a|\omega_x}$ for there to exist an explanation in terms of classical data $\Lambda^c_a[\omega_x]$. This noise is allowed to arise from any teleportation instrument, not necessarily classical one. 

The quantity $\mathcal{T}(\mathbb{\Lambda})$ is the (generalized) robustness of teleportation (RoT) and was introduced in \cite{Cavalcanti2017}. We highlight that for complete teleportation experiments the RoT is a function of the teleportation instrument $\mathbb{\Lambda}$ alone, and is independent of the specific set of states used $\{\omega_x\}$, and the data they produce $\{\sigma_{a|\omega_x}^\mathrm{B}\}$. We prove this important fact in the Appendix. 

\section{Results}
\subsection{Properties of Robustness of Teleportation}
Similarly to other robustness and weight-based measures \cite{Vidal1999,Napoli2016,Skrzypczyk2019,Theurer2019,Uola2019,Ducuara2019a}, the RoT has a number of useful properties which can be easily deduced from (\ref{eq:5}). Leaving the details to the Appendix, here we state the most important ones.

\noindent ($i$) It is \emph{faithful}, meaning that it vanishes if and only if teleportation instrument is classical, i.e:
    \begin{align}
        \mathcal{T}(\mathbb{\Lambda}) = 0 \iff \mathbb{\Lambda} \in \mathcal{F}.
    \end{align} 
($ii$) It is \emph{convex}, meaning that having access to teleportation instruments  $\mathbb{\Lambda}_1$ and $\mathbb{\Lambda}_2$ one cannot obtain a better one by using them probabilistically, i.e for $\Lam{}' = p\, \Lam{}_1 + (1-p)\,\Lam{}_2$ with 0 $\leq p \leq 1$, we have:
    \begin{align}
        \mathcal{T} \left( \Lam{}' \right) \leq p\, \mathcal{T}(\mathbb{\Lambda}_1) + (1-p)\,\mathcal{T}(\mathbb{\Lambda}_2).
    \end{align}
($iii$) It is \emph{monotonic} (non-increasing) under quantum and classical simulations. That is if $\Lam'{}$ can be simulated by $\Lam{}$ using a quantum or a classical simulation then
    \begin{align}
    \label{eq:18}
    \mathcal{T}(\mathbb{\Lambda}') \leq \mathcal{T}(\mathbb{\Lambda}).
    \end{align}
     A quantum simulation is one whereby there exist probability distributions $p_{\lambda}$, $p(b|a, \lambda)$ and channels $\Theta_{\lambda}$ and $\Omega_{\lambda}$ such that: 
    \begin{align}
        \label{eq:10}
        \Lambda_b' = \sum_{a, \lambda} p_{\lambda}\, p(b|a, \lambda)\, \Theta_{\lambda} \circ \Lambda_a \circ \Omega_{\lambda},
    \end{align}
    holds for all $b$. We denote the order induced by this type of simulation by $\Lam{}' \prec_q \Lam{}$. A classical simulation is one whereby there exist probability distributions $p(b|a)$ such that:
    \begin{align}
        \label{eq:010}
        \Lambda_b' = \sum_{a} p(b|a)\, \Lambda_a,
    \end{align}
    holds for all $b$ and is similarly denoted by $\Lam{}' \prec_c \Lam{}$. In the resource-theoretic approach one can think about these maps as free operations of the framework. The two  notions of simulation will each be seen to be relevant for one of the operational tasks introduced below.

\subsection{Operational Significance of Robustness of Teleportation}
Here we show that RoT can be viewed as the maximal achievable advantage when using quantum over classical resources in two unrelated operational tasks. Often it is illustrative to phrase such tasks in terms of games played between parties according to a pre-defined set of rules and scores. We follow this approach here and describe two operational tasks in terms of such games. 

\subsubsection{Teleportation of quantum correlations}
Consider a game played between a Verifier and Bob who tries to convince the Verifier about his ability to transfer correlations between two spatially separated labs. More explicitly, we consider the following scenario:
\begin{enumerate}
    \item The Verifier prepares an arbitrary bipartite state $\sigma^{\text{V'V}}$ and shares one part of this state with Bob.
    \item Bob inputs the state he received into a teleportation instrument $\Lam{}' = \{{\Lambda_b'}^{\text{V} \rightarrow \text{B}}\}_b$ which he can locally simulate using $\Lam{}$, obtaining measurement outcome $b$ and state $\sigma^{\text{V}'\text{B}}_{b} = (\mathcal{I}^{\text{V}'} \ot {\Lambda_b'}^{\text{V}\rightarrow \text{B}}) [\sigma^{\text{V}'\text{V}}]$. 
    \item Conditioned on the value of $b$ Bob applies locally a unitary correction $\mathcal{U}_b^{\text{B}}$ to his share of the state and returns the output state and outcome of the measurement to the Verifier.
    \item The Verifier assesses the quality of the teleportation instrument by checking the overlap between the joint state after correction $(\mathcal{I}^{\text{V}'}\ot\, \mathcal{U}_b^{\text{B}}) [\sigma^{\text{V}'\text{B}}_{b}]$ and a pre-defined set of target states $\{\xi_b^{\text{V}'\text{B}}\}$. If the teleported state is the same as the target state, then Bob receives a score $f(b) \geq 0$. 
\end{enumerate}
The game is fully specified by a tuple $\mathcal{G} = \{\sigma, \, \xi_b, f(b)\} $. The average score using the teleportation instrument $\Lam{}$ is given by:
\begin{align}
    \label{eq:11}
    q(\mathcal{G}, \mathbb{\Lambda})  = \max_{\substack{\Lam{}' \prec_q \Lam{} \\ \{\mathcal{U}_b\} \in \unitary }} \!\sum_{b} f(b) \tr \left[(\mathcal{I}\ot  \mathcal{U}_b \circ \Lambda_b')[\sigma] \cdot \xi_b \right],
\end{align} 
where the optimization ranges over all unitary corrections $\{\mathcal{U}_b\}$ and all teleportation instruments $\mathbb{\Lambda}'$ which can be quantum-simulated using $\mathbb{\Lambda}$, via (\ref{eq:10}).

In the Appendix we show that the maximal advantage which Bob can achieve using a teleportation instrument $\Lam{} \in \mathcal{R}$ over any classical instrument $\Lam^c \in \mathcal{F}$ is fully specified by the robustness of teleportation:
\begin{align}
    \max_{\mathcal{G}}\,  \frac{q(\mathcal{G}, \mathbb{\Lambda})}{ q^c(\mathcal{G})} = 1 + \mathcal{T}(\mathbb{\Lambda}),
\end{align}
where $q^c( \mathcal{G}) = \max_{\mathbb{\Lambda}^c \in \mathcal{F}} q(\mathcal{G}, \mathbb{\Lambda}^c)$  is the maximal score which can be achieved using classical resources in the same game (see Appendix for details). The proof technique is to (i) use (\ref{eq:5}) to show that  $ 1 + \mathcal{T}(\mathbb{\Lambda})$ is an upper bound on the advantage for all games $\mathcal{G}$; (ii) use duality theory of convex optimisation \cite{Boyd2004} to find the dual form of (\ref{eq:5}) and construct a  game $\mathcal{G}^*$ from the optimal dual variables that saturate the bound.  

\begin{figure}
    \centering
    \includegraphics[width=0.75\linewidth]{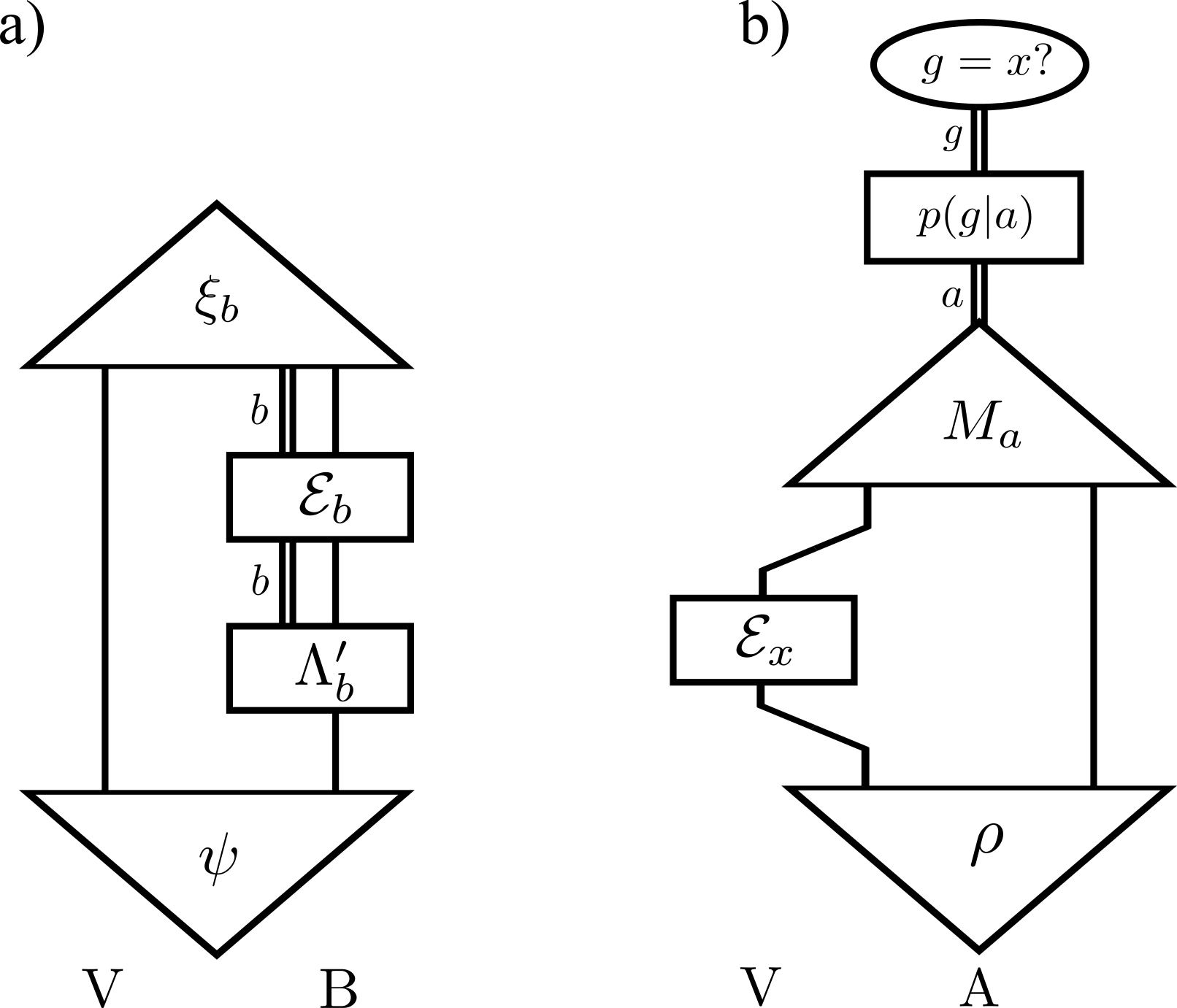}
    \caption{The two operational tasks. Fig. (a) presents teleportation of quantum correlations specified by $\mathcal{G} = \{\sigma, \xi_b, f(b)\}$, where $\sigma$ is the input state, $\{\xi_b\}_b$ are the target states and $f(b)$ is the score given when a correct target state is obtained. Bob is allowed to perform any local quantum simulation of his teleportation instrument $\Lam{}$, i.e. he has access to $\Lambda_b'$ of the form (\ref{eq:10}). Fig. (b) shows the task of subchannel discrimination with quantum side information which involves a set of subchannels to discriminate $\mathbb{E} = \{\mathcal{E}_x\}_x$ and uses quantum resources of the teleportation experiment (bipartite state and measurement).  }
    \label{fig:1}
\end{figure}

It is interesting that the average fidelity $\langle F \rangle$ can be viewed as the average score in this type of task for a particular game $\mathcal{G}$. To see this, consider a setting in which the Verifier provides a classically correlated state $\sigma^* = \frac{1}{n} \sum_x \dyad{x} \ot \dyad{\omega_x}$ and demands that the state returned by Bob is exactly the same for all $b$, that is $\xi_b^* = \sigma^* $.  For each transmission the Verifier will give Bob the same score $f^*(b) = n$.  This defines a game $\mathcal{G}^* = \{\sigma^*,\,  \xi_b^*, \, f^*(b) \}$, whose average score is: 
\begin{align}
    \label{eq:101}
    q(\mathcal{G}^*, \Lam{}) &=  \max_{\{U_a\} \in \unitary} \,\frac{1}{n}\, \sum_{a,x} p(a|x) \langle \omega_x | U_a \rho_{b|\omega_x}^{\text{B}} U_a^{\dagger}|\omega_x \rangle.
\end{align}
This is exactly the ordinary average fidelity (\ref{eq:65}). In interesting feature of this game is that Bob doesn't need to tell the Verifier which measurement result occurred. 

This provides insight into why not all entangled states are `useful' for teleportation. Since the average fidelity of teleportation corresponds to a game in which the Verifier asks Bob to transfer classical correlations, the fact that $\langle F \rangle$ cannot surpass the classical threshold for some entangled states only means that they cannot be used to transfer classical correlations better than the optimal classical state. However, if the verifier poses a more difficult talk where the correlations to be transferred are genuinely quantum, then all entangled states can outperform classical states for a specific choice of target states. Alternatively, one can view this task as a generalising from teleportation to entanglement swapping, in which both the input and target states can be both chosen arbitrarily.

\subsubsection{Subchannel discrimination with quantum side information and fixed measurement}
Let us now consider the task of subchannel discrimination, where the player is allowed to use a quantum memory to assist them, and only has the ability to perform a fixed entangled measurement. The task is specified by a collection of subchannels, $\mathbb{E} = \{\mathcal{E}_x\}$, which form an instrument. The resources of the player will be specified by $\mathcal{A} = \{\{M_a\}, \rho\}$, where $\{M_a\} \in \povm$ is a bipartite measurement and $\rho$ is the state of the quantum memory. We consider the following game set-up:
\begin{enumerate}
    \item Alice sends one half of the state $\rho^{\text{VA}}$ to the Verifier. 
    \item The Verifier applies a subchannel $\mathcal{E}_x^{\text{V}}$ from the instrument $\mathbb{E}$ to their share of $\rho^{\text{VA}}$, which results in $\rho_x^{\text{VA}} = (\mathcal{E}_x^{\text{V}} \ot \mathcal{I}^{\text{A}}) [\rho^{\text{VA}}]$ with probability $p(x|\rho) = \tr [\rho_x^{\text{VA}}]$. The Verifier then returns their share to Alice.   
    \item Alice uses the measurements $\{M_a^{\text{VA}}\}$ to identify which subchannel $\mathcal{E}_x^{\text{V}}$ was applied. Based on her measurement outcome $a$ she produces a guess $g$ according to $p(g|a)$.    
\end{enumerate}
The average probability of guessing which subchannel was applied when having access to $\rho$ and $\{M_a\}$, optimized over all post-processings $p(g|a)$ is given by:
\begin{align}
    \label{eq:100}
    \! p_{\text{succ}}(\mathbb{E}, \mathcal{A})\! =\! \max_{p(g|a)} \sum_{x,a,g} p(g|a) \tr[(\mathcal{E}_x \ot \mathcal{I})[\rho] \cdot M_a]  \delta_{g, x}.
\end{align}
We will compare this to the best success probability Alice could achieve if she had access to only classical resources. In particular, if either the memory used or the measurement performed is separable, then we will say that she uses a classical strategy $\mathcal{A}^c$ and denote the set of such strategies with $\mathcal{F}$. The (maximal) average guessing probability for such a classical strategy is given by $p_{\text{succ}}^c(\mathbb{E}) = \max_{\mathcal{A}^c \in \mathcal{F}} p_{\text{succ}}(\mathbb{E}, \mathcal{A}^c) $. It can be shown (see Appendix) that the optimal classical probability of guessing can be equivalently written as:
\begin{align}
    p_{\text{succ}}^c(\mathbb{E}) = \max_{\sigma} \max_{x} \, \,\, \tr \mathcal{E}_x(\sigma). 
\end{align}
In other words, the best classical strategy is to guess the most-likely outcome $x$ with the additional freedom to choose the probe state $\sigma$ which maximizes the guessing probability. 

In the Appendix we show that the maximal advantage offered by the strategy $\mathcal{A} = \{\{M_a\}, \rho\}$ over the best classical strategy is given by
\begin{align}
    \max_{\mathbb{E}}\,  \frac{p_{\text{succ}}(\mathbb{E}, \mathcal{A})}{ p_{\text{succ}}^c(\mathbb{E})} = 1 + \mathcal{T}(\mathbb{\Lambda}),
\end{align}
where $\Lam{}$ is the teleportation instrument formed by the measurement  $\{M_a\}$ and the state $\rho$. Thus, the maximal advantage is constant among all strategies $\mathcal{A}$ that lead to the same teleportation instrument $\Lam{}$. In the Appendix we show furthermore that $p_{\text{succ}}(\mathbb{E},\mathcal{A})$ in fact only depends on $\mathcal{A}$ through $\Lam{}$. 

The above reveals that the RoT fits into the program of robustness-based quantifiers and discrimination tasks, where the specific restrictions are on the resource state and resource measurement used to play the game. Interestingly, in \cite{Supic2019} the following relation between RoT and robustness of entanglement $R_{\,\text{E}}(\rho) := \min \{r \geq 0|\rho \leq (1+r)\sigma, \sigma \in \sep\}$ was shown:
\begin{align}
    \label{eq:rot_roe_c}
    \max_{\{M_a\} \in \povm} \mathcal{T}(\Lam{}) = R_{\,\text{E}}(\rho), 
\end{align}
This combined with our result provides a new operational meaning for the (generalized) robustness of entanglement: it quantifies the advantage entangled states offer when acting as quantum memories in local subchannel discrimination, i.e.:
\begin{align}
    \max_{\mathbb{E}}\,\max_{\{M_a\} \in \povm}\,\frac{p_{\text{succ}}(\mathbb{E}, \{M_a\}, \rho)}{ p_{\text{succ}}^c(\mathbb{E})} = 1 + R_{\,\text{E}}(\rho),
\end{align}
In other words, every entangled state can act as a useful quantum memory in \emph{local} subchannel discrimination.

\subsubsection{Complete sets of monotones for teleportation simulation}
The average score (\ref{eq:11}) and average guessing probability (\ref{eq:100}) are also important as they provide complete characterisations for the two notions of teleportation simulation introduced in \eqref{eq:10} and \eqref{eq:010}. In particular, in the Appendix we show that $\Lam{}$ can quantum-simulate $\Lam{}'$, $\Lam{} \succ_q \Lam{}'$ if and only if
\begin{align}
    q(\mathcal{G}, \Lam{}) \geq q(\mathcal{G}, \Lam{}') \text{ for all games $\mathcal{G}$}
\end{align}
Similarly, $\Lam{}$ can classically simulate $\Lam{}'$, $\Lam{} \succ_c \Lam{}'$ if and only if:
\begin{align}
    p_{\text{succ}}(\mathbb{E}, \Lam{}) \geq p_{\text{succ}}(\mathbb{E}, \Lam{}') \text{ for all games $\mathbb{E}$},
\end{align}
 This means that both $q(\mathcal{G}, \Lam{})$ and $ p_{\text{succ}}(\mathbb{E}, \Lam{})$ constitute ``complete set of monotones'', the former for the partial order of quantum-simulation, and the latter for classical-simulation. 

\section{Conclusions}
We have analysed a robustness-based quantifier of teleportation and shown that it has operational significance in two unrelated directions. On the one hand it quantifies the advantage that a given teleportation instrument offers for the task of teleporting quantum correlations. On the other hand, it also quantifies the advantage offered by a fixed entangled state and fixed entangled measurement in the task of subchannel discrimination with side information. 

We showed that the first task is a natural generalisation of the standard task used for benchmarking the quality of a teleportation set-up (the average fidelity of teleportation), and thus provides an answer to the question of in what sense is every state useful for teleportation: Every state has the ability to teleport quantum correlations strictly better than can be achieved by any classical teleportation scheme. 

We finally showed that the two tasks which give operational meaning to the robustness of teleportation also form complete sets of monotones, which fully characterise two natural notions of simulation that arise for teleportation, one purely classical, and the other quantum.  
\newline
\begin{acknowledgements}
We would like to thank Andr\'es Ducuara and Tom Purves for helpful and inspiring discussions.  PLB acknowledges support from the UK EPSRC (grant no. EP/R00644X/1). PS acknowledges support from a Royal Society URF (UHQT).
\end{acknowledgements}

\bibliographystyle{apsrev4-1}
\bibliography{apssamp}

\providecommand{\noopsort}[1]{}\providecommand{\singleletter}[1]{#1}%
\begin{thebibliography}{69}%
\makeatletter
\providecommand \@ifxundefined [1]{%
 \@ifx{#1\undefined}
}%
\providecommand \@ifnum [1]{%
 \ifnum #1\expandafter \@firstoftwo
 \else \expandafter \@secondoftwo
 \fi
}%
\providecommand \@ifx [1]{%
 \ifx #1\expandafter \@firstoftwo
 \else \expandafter \@secondoftwo
 \fi
}%
\providecommand \natexlab [1]{#1}%
\providecommand \enquote  [1]{``#1''}%
\providecommand \bibnamefont  [1]{#1}%
\providecommand \bibfnamefont [1]{#1}%
\providecommand \citenamefont [1]{#1}%
\providecommand \href@noop [0]{\@secondoftwo}%
\providecommand \href [0]{\begingroup \@sanitize@url \@href}%
\providecommand \@href[1]{\@@startlink{#1}\@@href}%
\providecommand \@@href[1]{\endgroup#1\@@endlink}%
\providecommand \@sanitize@url [0]{\catcode `\\12\catcode `\$12\catcode
  `\&12\catcode `\#12\catcode `\^12\catcode `\_12\catcode `\%12\relax}%
\providecommand \@@startlink[1]{}%
\providecommand \@@endlink[0]{}%
\providecommand \url  [0]{\begingroup\@sanitize@url \@url }%
\providecommand \@url [1]{\endgroup\@href {#1}{\urlprefix }}%
\providecommand \urlprefix  [0]{URL }%
\providecommand \Eprint [0]{\href }%
\providecommand \doibase [0]{http://dx.doi.org/}%
\providecommand \selectlanguage [0]{\@gobble}%
\providecommand \bibinfo  [0]{\@secondoftwo}%
\providecommand \bibfield  [0]{\@secondoftwo}%
\providecommand \translation [1]{[#1]}%
\providecommand \BibitemOpen [0]{}%
\providecommand \bibitemStop [0]{}%
\providecommand \bibitemNoStop [0]{.\EOS\space}%
\providecommand \EOS [0]{\spacefactor3000\relax}%
\providecommand \BibitemShut  [1]{\csname bibitem#1\endcsname}%
\let\auto@bib@innerbib\@empty
\bibitem [{\citenamefont {Bennett}\ \emph {et~al.}(1993)\citenamefont
  {Bennett}, \citenamefont {Brassard}, \citenamefont {Cr\'epeau}, \citenamefont
  {Jozsa}, \citenamefont {Peres},\ and\ \citenamefont {Wootters}}]{Bennet1993}%
  \BibitemOpen
  \bibfield  {author} {\bibinfo {author} {\bibfnamefont {C.~H.}\ \bibnamefont
  {Bennett}}, \bibinfo {author} {\bibfnamefont {G.}~\bibnamefont {Brassard}},
  \bibinfo {author} {\bibfnamefont {C.}~\bibnamefont {Cr\'epeau}}, \bibinfo
  {author} {\bibfnamefont {R.}~\bibnamefont {Jozsa}}, \bibinfo {author}
  {\bibfnamefont {A.}~\bibnamefont {Peres}}, \ and\ \bibinfo {author}
  {\bibfnamefont {W.~K.}\ \bibnamefont {Wootters}},\ }\href {\doibase
  10.1103/PhysRevLett.70.1895} {\bibfield  {journal} {\bibinfo  {journal}
  {Phys. Rev. Lett.}\ }\textbf {\bibinfo {volume} {70}},\ \bibinfo {pages}
  {1895} (\bibinfo {year} {1993})}\BibitemShut {NoStop}%
\bibitem [{\citenamefont {Bouwmeester}\ \emph {et~al.}(1997)\citenamefont
  {Bouwmeester}, \citenamefont {Pan}, \citenamefont {Mattle}, \citenamefont
  {Eibl}, \citenamefont {Weinfurter},\ and\ \citenamefont
  {Zeilinger}}]{Bouwmeester1997}%
  \BibitemOpen
  \bibfield  {author} {\bibinfo {author} {\bibfnamefont {D.}~\bibnamefont
  {Bouwmeester}}, \bibinfo {author} {\bibfnamefont {J.-W.}\ \bibnamefont
  {Pan}}, \bibinfo {author} {\bibfnamefont {K.}~\bibnamefont {Mattle}},
  \bibinfo {author} {\bibfnamefont {M.}~\bibnamefont {Eibl}}, \bibinfo {author}
  {\bibfnamefont {H.}~\bibnamefont {Weinfurter}}, \ and\ \bibinfo {author}
  {\bibfnamefont {A.}~\bibnamefont {Zeilinger}},\ }\href {\doibase
  10.1038/37539} {\bibfield  {journal} {\bibinfo  {journal} {Nature}\ }\textbf
  {\bibinfo {volume} {390}},\ \bibinfo {pages} {575} (\bibinfo {year}
  {1997})}\BibitemShut {NoStop}%
\bibitem [{\citenamefont {Boschi}\ \emph {et~al.}(1998)\citenamefont {Boschi},
  \citenamefont {Branca}, \citenamefont {De~Martini}, \citenamefont {Hardy},\
  and\ \citenamefont {Popescu}}]{Boschi1998}%
  \BibitemOpen
  \bibfield  {author} {\bibinfo {author} {\bibfnamefont {D.}~\bibnamefont
  {Boschi}}, \bibinfo {author} {\bibfnamefont {S.}~\bibnamefont {Branca}},
  \bibinfo {author} {\bibfnamefont {F.}~\bibnamefont {De~Martini}}, \bibinfo
  {author} {\bibfnamefont {L.}~\bibnamefont {Hardy}}, \ and\ \bibinfo {author}
  {\bibfnamefont {S.}~\bibnamefont {Popescu}},\ }\href {\doibase
  10.1103/PhysRevLett.80.1121} {\bibfield  {journal} {\bibinfo  {journal}
  {Phys. Rev. Lett.}\ }\textbf {\bibinfo {volume} {80}},\ \bibinfo {pages}
  {1121} (\bibinfo {year} {1998})}\BibitemShut {NoStop}%
\bibitem [{\citenamefont {Furusawa}\ \emph {et~al.}(1998)\citenamefont
  {Furusawa}, \citenamefont {S{\o}rensen}, \citenamefont {Braunstein},
  \citenamefont {Fuchs}, \citenamefont {Kimble},\ and\ \citenamefont
  {Polzik}}]{Furusawa1998}%
  \BibitemOpen
  \bibfield  {author} {\bibinfo {author} {\bibfnamefont {A.}~\bibnamefont
  {Furusawa}}, \bibinfo {author} {\bibfnamefont {J.~L.}\ \bibnamefont
  {S{\o}rensen}}, \bibinfo {author} {\bibfnamefont {S.~L.}\ \bibnamefont
  {Braunstein}}, \bibinfo {author} {\bibfnamefont {C.~A.}\ \bibnamefont
  {Fuchs}}, \bibinfo {author} {\bibfnamefont {H.~J.}\ \bibnamefont {Kimble}}, \
  and\ \bibinfo {author} {\bibfnamefont {E.~S.}\ \bibnamefont {Polzik}},\
  }\href {\doibase 10.1126/science.282.5389.706} {\bibfield  {journal}
  {\bibinfo  {journal} {Science}\ }\textbf {\bibinfo {volume} {282}},\ \bibinfo
  {pages} {706} (\bibinfo {year} {1998})}\BibitemShut {NoStop}%
\bibitem [{\citenamefont {Bao}\ \emph {et~al.}(2012)\citenamefont {Bao},
  \citenamefont {Xu}, \citenamefont {Li}, \citenamefont {Yuan}, \citenamefont
  {Lu},\ and\ \citenamefont {Pan}}]{Bao2012}%
  \BibitemOpen
  \bibfield  {author} {\bibinfo {author} {\bibfnamefont {X.-H.}\ \bibnamefont
  {Bao}}, \bibinfo {author} {\bibfnamefont {X.-F.}\ \bibnamefont {Xu}},
  \bibinfo {author} {\bibfnamefont {C.-M.}\ \bibnamefont {Li}}, \bibinfo
  {author} {\bibfnamefont {Z.-S.}\ \bibnamefont {Yuan}}, \bibinfo {author}
  {\bibfnamefont {C.-Y.}\ \bibnamefont {Lu}}, \ and\ \bibinfo {author}
  {\bibfnamefont {J.-W.}\ \bibnamefont {Pan}},\ }\href {\doibase
  10.1073/pnas.1207329109} {\bibfield  {journal} {\bibinfo  {journal} {Proc.
  Natl. Acad. Sci.}\ }\textbf {\bibinfo {volume} {109}},\ \bibinfo {pages}
  {20347} (\bibinfo {year} {2012})}\BibitemShut {NoStop}%
\bibitem [{\citenamefont {Leuenberger}\ \emph {et~al.}(2005)\citenamefont
  {Leuenberger}, \citenamefont {Flatt\'e},\ and\ \citenamefont
  {Awschalom}}]{Leuenberger2005}%
  \BibitemOpen
  \bibfield  {author} {\bibinfo {author} {\bibfnamefont {M.~N.}\ \bibnamefont
  {Leuenberger}}, \bibinfo {author} {\bibfnamefont {M.~E.}\ \bibnamefont
  {Flatt\'e}}, \ and\ \bibinfo {author} {\bibfnamefont {D.~D.}\ \bibnamefont
  {Awschalom}},\ }\href {\doibase 10.1103/PhysRevLett.94.107401} {\bibfield
  {journal} {\bibinfo  {journal} {Phys. Rev. Lett.}\ }\textbf {\bibinfo
  {volume} {94}},\ \bibinfo {pages} {107401} (\bibinfo {year}
  {2005})}\BibitemShut {NoStop}%
\bibitem [{\citenamefont {Pirandola}\ \emph {et~al.}(2015)\citenamefont
  {Pirandola}, \citenamefont {Eisert}, \citenamefont {Weedbrook}, \citenamefont
  {Furusawa},\ and\ \citenamefont {Braunstein}}]{Pirandola2015}%
  \BibitemOpen
  \bibfield  {author} {\bibinfo {author} {\bibfnamefont {S.}~\bibnamefont
  {Pirandola}}, \bibinfo {author} {\bibfnamefont {J.}~\bibnamefont {Eisert}},
  \bibinfo {author} {\bibfnamefont {C.}~\bibnamefont {Weedbrook}}, \bibinfo
  {author} {\bibfnamefont {A.}~\bibnamefont {Furusawa}}, \ and\ \bibinfo
  {author} {\bibfnamefont {S.~L.}\ \bibnamefont {Braunstein}},\ }\href
  {https://doi.org/10.1038/nphoton.2015.154} {\bibfield  {journal} {\bibinfo
  {journal} {Nature Photonics}\ }\textbf {\bibinfo {volume} {9}} (\bibinfo
  {year} {2015})}\BibitemShut {NoStop}%
\bibitem [{\citenamefont {Vaidman}(1994)}]{Vaidman1994}%
  \BibitemOpen
  \bibfield  {author} {\bibinfo {author} {\bibfnamefont {L.}~\bibnamefont
  {Vaidman}},\ }\href {\doibase 10.1103/PhysRevA.49.1473} {\bibfield  {journal}
  {\bibinfo  {journal} {Phys. Rev. A}\ }\textbf {\bibinfo {volume} {49}},\
  \bibinfo {pages} {1473} (\bibinfo {year} {1994})}\BibitemShut {NoStop}%
\bibitem [{\citenamefont {Sherson}\ \emph {et~al.}(2006)\citenamefont
  {Sherson}, \citenamefont {Krauter}, \citenamefont {Olsson}, \citenamefont
  {Julsgaard}, \citenamefont {Hammerer}, \citenamefont {Cirac},\ and\
  \citenamefont {Polzik}}]{Sherson2006}%
  \BibitemOpen
  \bibfield  {author} {\bibinfo {author} {\bibfnamefont {J.~F.}\ \bibnamefont
  {Sherson}}, \bibinfo {author} {\bibfnamefont {H.}~\bibnamefont {Krauter}},
  \bibinfo {author} {\bibfnamefont {R.~K.}\ \bibnamefont {Olsson}}, \bibinfo
  {author} {\bibfnamefont {B.}~\bibnamefont {Julsgaard}}, \bibinfo {author}
  {\bibfnamefont {K.}~\bibnamefont {Hammerer}}, \bibinfo {author}
  {\bibfnamefont {I.}~\bibnamefont {Cirac}}, \ and\ \bibinfo {author}
  {\bibfnamefont {E.~S.}\ \bibnamefont {Polzik}},\ }\href {\doibase
  10.1038/nature05136} {\bibfield  {journal} {\bibinfo  {journal} {Nature}\
  }\textbf {\bibinfo {volume} {443}},\ \bibinfo {pages} {557} (\bibinfo {year}
  {2006})}\BibitemShut {NoStop}%
\bibitem [{\citenamefont {Briegel}\ \emph {et~al.}(1998)\citenamefont
  {Briegel}, \citenamefont {D\"ur}, \citenamefont {Cirac},\ and\ \citenamefont
  {Zoller}}]{Briegel1998}%
  \BibitemOpen
  \bibfield  {author} {\bibinfo {author} {\bibfnamefont {H.-J.}\ \bibnamefont
  {Briegel}}, \bibinfo {author} {\bibfnamefont {W.}~\bibnamefont {D\"ur}},
  \bibinfo {author} {\bibfnamefont {J.~I.}\ \bibnamefont {Cirac}}, \ and\
  \bibinfo {author} {\bibfnamefont {P.}~\bibnamefont {Zoller}},\ }\href
  {\doibase 10.1103/PhysRevLett.81.5932} {\bibfield  {journal} {\bibinfo
  {journal} {Phys. Rev. Lett.}\ }\textbf {\bibinfo {volume} {81}},\ \bibinfo
  {pages} {5932} (\bibinfo {year} {1998})}\BibitemShut {NoStop}%
\bibitem [{\citenamefont {Hasegawa}\ \emph {et~al.}(2019)\citenamefont
  {Hasegawa}, \citenamefont {Ikuta}, \citenamefont {Matsuda}, \citenamefont
  {Tamaki}, \citenamefont {Lo}, \citenamefont {Yamamoto}, \citenamefont
  {Azuma},\ and\ \citenamefont {Imoto}}]{Hasegawa2019}%
  \BibitemOpen
  \bibfield  {author} {\bibinfo {author} {\bibfnamefont {Y.}~\bibnamefont
  {Hasegawa}}, \bibinfo {author} {\bibfnamefont {R.}~\bibnamefont {Ikuta}},
  \bibinfo {author} {\bibfnamefont {N.}~\bibnamefont {Matsuda}}, \bibinfo
  {author} {\bibfnamefont {K.}~\bibnamefont {Tamaki}}, \bibinfo {author}
  {\bibfnamefont {H.-K.}\ \bibnamefont {Lo}}, \bibinfo {author} {\bibfnamefont
  {T.}~\bibnamefont {Yamamoto}}, \bibinfo {author} {\bibfnamefont
  {K.}~\bibnamefont {Azuma}}, \ and\ \bibinfo {author} {\bibfnamefont
  {N.}~\bibnamefont {Imoto}},\ }\href {\doibase 10.1038/s41467-018-08099-5}
  {\bibfield  {journal} {\bibinfo  {journal} {Nat. Commun.}\ }\textbf {\bibinfo
  {volume} {10}},\ \bibinfo {pages} {378} (\bibinfo {year} {2019})}\BibitemShut
  {NoStop}%
\bibitem [{\citenamefont {Gottesman}\ and\ \citenamefont
  {Chuang}(1999)}]{Gottesman1999}%
  \BibitemOpen
  \bibfield  {author} {\bibinfo {author} {\bibfnamefont {D.}~\bibnamefont
  {Gottesman}}\ and\ \bibinfo {author} {\bibfnamefont {I.~L.}\ \bibnamefont
  {Chuang}},\ }\href {\doibase 10.1038/46503} {\bibfield  {journal} {\bibinfo
  {journal} {Nature}\ }\textbf {\bibinfo {volume} {402}},\ \bibinfo {pages}
  {390} (\bibinfo {year} {1999})}\BibitemShut {NoStop}%
\bibitem [{\citenamefont {Kimble}(2008)}]{Kimble2008}%
  \BibitemOpen
  \bibfield  {author} {\bibinfo {author} {\bibfnamefont {H.~J.}\ \bibnamefont
  {Kimble}},\ }\href {https://doi.org/10.1038/nature07127} {\bibfield
  {journal} {\bibinfo  {journal} {Nature}\ }\textbf {\bibinfo {volume} {453}},\
  \bibinfo {pages} {1023} (\bibinfo {year} {2008})}\BibitemShut {NoStop}%
\bibitem [{\citenamefont {Popescu}(1994)}]{Popescu1994}%
  \BibitemOpen
  \bibfield  {author} {\bibinfo {author} {\bibfnamefont {S.}~\bibnamefont
  {Popescu}},\ }\href {\doibase 10.1103/PhysRevLett.72.797} {\bibfield
  {journal} {\bibinfo  {journal} {Phys. Rev. Lett.}\ }\textbf {\bibinfo
  {volume} {72}},\ \bibinfo {pages} {797} (\bibinfo {year} {1994})}\BibitemShut
  {NoStop}%
\bibitem [{\citenamefont {Horodecki}\ \emph {et~al.}(1999)\citenamefont
  {Horodecki}, \citenamefont {Horodecki},\ and\ \citenamefont
  {Horodecki}}]{Horodecki1999}%
  \BibitemOpen
  \bibfield  {author} {\bibinfo {author} {\bibfnamefont {M.}~\bibnamefont
  {Horodecki}}, \bibinfo {author} {\bibfnamefont {P.}~\bibnamefont
  {Horodecki}}, \ and\ \bibinfo {author} {\bibfnamefont {R.}~\bibnamefont
  {Horodecki}},\ }\href {\doibase 10.1103/PhysRevA.60.1888} {\bibfield
  {journal} {\bibinfo  {journal} {Phys. Rev. A}\ }\textbf {\bibinfo {volume}
  {60}},\ \bibinfo {pages} {1888} (\bibinfo {year} {1999})}\BibitemShut
  {NoStop}%
\bibitem [{\citenamefont {Linden}\ and\ \citenamefont
  {Popescu}(1999)}]{Linden1999}%
  \BibitemOpen
  \bibfield  {author} {\bibinfo {author} {\bibfnamefont {N.}~\bibnamefont
  {Linden}}\ and\ \bibinfo {author} {\bibfnamefont {S.}~\bibnamefont
  {Popescu}},\ }\href {\doibase 10.1103/PhysRevA.59.137} {\bibfield  {journal}
  {\bibinfo  {journal} {Phys. Rev. A}\ }\textbf {\bibinfo {volume} {59}},\
  \bibinfo {pages} {137} (\bibinfo {year} {1999})}\BibitemShut {NoStop}%
\bibitem [{\citenamefont {Olmschenk}\ \emph {et~al.}(2009)\citenamefont
  {Olmschenk}, \citenamefont {Matsukevich}, \citenamefont {Maunz},
  \citenamefont {Hayes}, \citenamefont {Duan},\ and\ \citenamefont
  {Monroe}}]{Olmschenk2009}%
  \BibitemOpen
  \bibfield  {author} {\bibinfo {author} {\bibfnamefont {S.}~\bibnamefont
  {Olmschenk}}, \bibinfo {author} {\bibfnamefont {D.}~\bibnamefont
  {Matsukevich}}, \bibinfo {author} {\bibfnamefont {P.}~\bibnamefont {Maunz}},
  \bibinfo {author} {\bibfnamefont {D.}~\bibnamefont {Hayes}}, \bibinfo
  {author} {\bibfnamefont {L.-M.}\ \bibnamefont {Duan}}, \ and\ \bibinfo
  {author} {\bibfnamefont {C.}~\bibnamefont {Monroe}},\ }\href@noop {}
  {\bibfield  {journal} {\bibinfo  {journal} {Science}\ }\textbf {\bibinfo
  {volume} {323}},\ \bibinfo {pages} {486} (\bibinfo {year}
  {2009})}\BibitemShut {NoStop}%
\bibitem [{\citenamefont {Bennett}\ \emph {et~al.}(1999)\citenamefont
  {Bennett}, \citenamefont {DiVincenzo}, \citenamefont {Mor}, \citenamefont
  {Shor}, \citenamefont {Smolin},\ and\ \citenamefont {Terhal}}]{Bennet1999}%
  \BibitemOpen
  \bibfield  {author} {\bibinfo {author} {\bibfnamefont {C.~H.}\ \bibnamefont
  {Bennett}}, \bibinfo {author} {\bibfnamefont {D.~P.}\ \bibnamefont
  {DiVincenzo}}, \bibinfo {author} {\bibfnamefont {T.}~\bibnamefont {Mor}},
  \bibinfo {author} {\bibfnamefont {P.~W.}\ \bibnamefont {Shor}}, \bibinfo
  {author} {\bibfnamefont {J.~A.}\ \bibnamefont {Smolin}}, \ and\ \bibinfo
  {author} {\bibfnamefont {B.~M.}\ \bibnamefont {Terhal}},\ }\href {\doibase
  10.1103/PhysRevLett.82.5385} {\bibfield  {journal} {\bibinfo  {journal}
  {Phys. Rev. Lett.}\ }\textbf {\bibinfo {volume} {82}},\ \bibinfo {pages}
  {5385} (\bibinfo {year} {1999})}\BibitemShut {NoStop}%
\bibitem [{\citenamefont {Horodecki}\ \emph
  {et~al.}(2003{\natexlab{a}})\citenamefont {Horodecki}, \citenamefont
  {Smolin}, \citenamefont {Terhal},\ and\ \citenamefont
  {Thapliyal}}]{HorodeckiP2003}%
  \BibitemOpen
  \bibfield  {author} {\bibinfo {author} {\bibfnamefont {P.}~\bibnamefont
  {Horodecki}}, \bibinfo {author} {\bibfnamefont {J.~A.}\ \bibnamefont
  {Smolin}}, \bibinfo {author} {\bibfnamefont {B.~M.}\ \bibnamefont {Terhal}},
  \ and\ \bibinfo {author} {\bibfnamefont {A.~V.}\ \bibnamefont {Thapliyal}},\
  }\href {\doibase https://doi.org/10.1016/S0304-3975(01)00376-0} {\bibfield
  {journal} {\bibinfo  {journal} {Theoretical Computer Science}\ }\textbf
  {\bibinfo {volume} {292}},\ \bibinfo {pages} {589 } (\bibinfo {year}
  {2003}{\natexlab{a}})},\ \bibinfo {note} {algorithms in Quantum Information
  Prcoessing}\BibitemShut {NoStop}%
\bibitem [{\citenamefont {Horodecki}\ \emph {et~al.}(1998)\citenamefont
  {Horodecki}, \citenamefont {Horodecki},\ and\ \citenamefont
  {Horodecki}}]{Horodecki1998}%
  \BibitemOpen
  \bibfield  {author} {\bibinfo {author} {\bibfnamefont {M.}~\bibnamefont
  {Horodecki}}, \bibinfo {author} {\bibfnamefont {P.}~\bibnamefont
  {Horodecki}}, \ and\ \bibinfo {author} {\bibfnamefont {R.}~\bibnamefont
  {Horodecki}},\ }\href {\doibase 10.1103/PhysRevLett.80.5239} {\bibfield
  {journal} {\bibinfo  {journal} {Phys. Rev. Lett.}\ }\textbf {\bibinfo
  {volume} {80}},\ \bibinfo {pages} {5239} (\bibinfo {year}
  {1998})}\BibitemShut {NoStop}%
\bibitem [{\citenamefont {Cavalcanti}\ \emph {et~al.}(2017)\citenamefont
  {Cavalcanti}, \citenamefont {Skrzypczyk},\ and\ \citenamefont {\ifmmode
  \check{S}\else \v{S}\fi{}upi\ifmmode~\acute{c}\else
  \'{c}\fi{}}}]{Cavalcanti2017}%
  \BibitemOpen
  \bibfield  {author} {\bibinfo {author} {\bibfnamefont {D.}~\bibnamefont
  {Cavalcanti}}, \bibinfo {author} {\bibfnamefont {P.}~\bibnamefont
  {Skrzypczyk}}, \ and\ \bibinfo {author} {\bibfnamefont {I.}~\bibnamefont
  {\ifmmode \check{S}\else \v{S}\fi{}upi\ifmmode~\acute{c}\else \'{c}\fi{}}},\
  }\href {\doibase 10.1103/PhysRevLett.119.110501} {\bibfield  {journal}
  {\bibinfo  {journal} {Phys. Rev. Lett.}\ }\textbf {\bibinfo {volume} {119}},\
  \bibinfo {pages} {110501} (\bibinfo {year} {2017})}\BibitemShut {NoStop}%
\bibitem [{\citenamefont {\ifmmode \check{S}\else
  \v{S}\fi{}upi\ifmmode~\acute{c}\else \'{c}\fi{}}\ \emph
  {et~al.}(2019)\citenamefont {\ifmmode \check{S}\else
  \v{S}\fi{}upi\ifmmode~\acute{c}\else \'{c}\fi{}}, \citenamefont
  {Skrzypczyk},\ and\ \citenamefont {Cavalcanti}}]{Supic2019}%
  \BibitemOpen
  \bibfield  {author} {\bibinfo {author} {\bibfnamefont {I.}~\bibnamefont
  {\ifmmode \check{S}\else \v{S}\fi{}upi\ifmmode~\acute{c}\else \'{c}\fi{}}},
  \bibinfo {author} {\bibfnamefont {P.}~\bibnamefont {Skrzypczyk}}, \ and\
  \bibinfo {author} {\bibfnamefont {D.}~\bibnamefont {Cavalcanti}},\ }\href
  {\doibase 10.1103/PhysRevA.99.032334} {\bibfield  {journal} {\bibinfo
  {journal} {Phys. Rev. A}\ }\textbf {\bibinfo {volume} {99}},\ \bibinfo
  {pages} {032334} (\bibinfo {year} {2019})}\BibitemShut {NoStop}%
\bibitem [{\citenamefont {\ifmmode~\dot{Z}\else \.{Z}\fi{}ukowski}\ \emph
  {et~al.}(1993)\citenamefont {\ifmmode~\dot{Z}\else \.{Z}\fi{}ukowski},
  \citenamefont {Zeilinger}, \citenamefont {Horne},\ and\ \citenamefont
  {Ekert}}]{Zukowski1993}%
  \BibitemOpen
  \bibfield  {author} {\bibinfo {author} {\bibfnamefont {M.}~\bibnamefont
  {\ifmmode~\dot{Z}\else \.{Z}\fi{}ukowski}}, \bibinfo {author} {\bibfnamefont
  {A.}~\bibnamefont {Zeilinger}}, \bibinfo {author} {\bibfnamefont {M.~A.}\
  \bibnamefont {Horne}}, \ and\ \bibinfo {author} {\bibfnamefont {A.~K.}\
  \bibnamefont {Ekert}},\ }\href {\doibase 10.1103/PhysRevLett.71.4287}
  {\bibfield  {journal} {\bibinfo  {journal} {Phys. Rev. Lett.}\ }\textbf
  {\bibinfo {volume} {71}},\ \bibinfo {pages} {4287} (\bibinfo {year}
  {1993})}\BibitemShut {NoStop}%
\bibitem [{\citenamefont {Pan}\ \emph {et~al.}(1998)\citenamefont {Pan},
  \citenamefont {Bouwmeester}, \citenamefont {Weinfurter},\ and\ \citenamefont
  {Zeilinger}}]{Pan1998}%
  \BibitemOpen
  \bibfield  {author} {\bibinfo {author} {\bibfnamefont {J.-W.}\ \bibnamefont
  {Pan}}, \bibinfo {author} {\bibfnamefont {D.}~\bibnamefont {Bouwmeester}},
  \bibinfo {author} {\bibfnamefont {H.}~\bibnamefont {Weinfurter}}, \ and\
  \bibinfo {author} {\bibfnamefont {A.}~\bibnamefont {Zeilinger}},\ }\href
  {\doibase 10.1103/PhysRevLett.80.3891} {\bibfield  {journal} {\bibinfo
  {journal} {Phys. Rev. Lett.}\ }\textbf {\bibinfo {volume} {80}},\ \bibinfo
  {pages} {3891} (\bibinfo {year} {1998})}\BibitemShut {NoStop}%
\bibitem [{\citenamefont {Kitaev}(1997)}]{Kitaev1997}%
  \BibitemOpen
  \bibfield  {author} {\bibinfo {author} {\bibfnamefont {A.~Y.}\ \bibnamefont
  {Kitaev}},\ }\href {\doibase 10.1070/rm1997v052n06abeh002155} {\bibfield
  {journal} {\bibinfo  {journal} {Russian Mathematical Surveys}\ }\textbf
  {\bibinfo {volume} {52}},\ \bibinfo {pages} {1191} (\bibinfo {year}
  {1997})}\BibitemShut {NoStop}%
\bibitem [{\citenamefont {Ac\'{\i}n}(2001)}]{Acin20011}%
  \BibitemOpen
  \bibfield  {author} {\bibinfo {author} {\bibfnamefont {A.}~\bibnamefont
  {Ac\'{\i}n}},\ }\href {\doibase 10.1103/PhysRevLett.87.177901} {\bibfield
  {journal} {\bibinfo  {journal} {Phys. Rev. Lett.}\ }\textbf {\bibinfo
  {volume} {87}},\ \bibinfo {pages} {177901} (\bibinfo {year}
  {2001})}\BibitemShut {NoStop}%
\bibitem [{\citenamefont {Childs}\ \emph {et~al.}(2000)\citenamefont {Childs},
  \citenamefont {Preskill},\ and\ \citenamefont {Renes}}]{Childs2000}%
  \BibitemOpen
  \bibfield  {author} {\bibinfo {author} {\bibfnamefont {A.~M.}\ \bibnamefont
  {Childs}}, \bibinfo {author} {\bibfnamefont {J.}~\bibnamefont {Preskill}}, \
  and\ \bibinfo {author} {\bibfnamefont {J.}~\bibnamefont {Renes}},\ }\href
  {\doibase 10.1080/09500340008244034} {\bibfield  {journal} {\bibinfo
  {journal} {Journal of Modern Optics}\ }\textbf {\bibinfo {volume} {47}},\
  \bibinfo {pages} {155} (\bibinfo {year} {2000})},\ \Eprint
  {http://arxiv.org/abs/https://www.tandfonline.com/doi/pdf/10.1080/09500340008244034}
  {https://www.tandfonline.com/doi/pdf/10.1080/09500340008244034} \BibitemShut
  {NoStop}%
\bibitem [{\citenamefont {Vidal}\ and\ \citenamefont
  {Tarrach}(1999)}]{Vidal1999}%
  \BibitemOpen
  \bibfield  {author} {\bibinfo {author} {\bibfnamefont {G.}~\bibnamefont
  {Vidal}}\ and\ \bibinfo {author} {\bibfnamefont {R.}~\bibnamefont
  {Tarrach}},\ }\href@noop {} {\bibfield  {journal} {\bibinfo  {journal}
  {Physical Review A}\ }\textbf {\bibinfo {volume} {59}},\ \bibinfo {pages}
  {141} (\bibinfo {year} {1999})}\BibitemShut {NoStop}%
\bibitem [{\citenamefont {Bae}\ \emph {et~al.}(2019)\citenamefont {Bae},
  \citenamefont {Chru\ifmmode \acute{s}\else
  \'{s}\fi{}ci\ifmmode~\acute{n}\else \'{n}\fi{}ski},\ and\ \citenamefont
  {Piani}}]{Bae2019}%
  \BibitemOpen
  \bibfield  {author} {\bibinfo {author} {\bibfnamefont {J.}~\bibnamefont
  {Bae}}, \bibinfo {author} {\bibfnamefont {D.}~\bibnamefont {Chru\ifmmode
  \acute{s}\else \'{s}\fi{}ci\ifmmode~\acute{n}\else \'{n}\fi{}ski}}, \ and\
  \bibinfo {author} {\bibfnamefont {M.}~\bibnamefont {Piani}},\ }\href
  {\doibase 10.1103/PhysRevLett.122.140404} {\bibfield  {journal} {\bibinfo
  {journal} {Phys. Rev. Lett.}\ }\textbf {\bibinfo {volume} {122}},\ \bibinfo
  {pages} {140404} (\bibinfo {year} {2019})}\BibitemShut {NoStop}%
\bibitem [{\citenamefont {Takagi}\ and\ \citenamefont
  {Regula}(2019)}]{Takagi2019}%
  \BibitemOpen
  \bibfield  {author} {\bibinfo {author} {\bibfnamefont {R.}~\bibnamefont
  {Takagi}}\ and\ \bibinfo {author} {\bibfnamefont {B.}~\bibnamefont
  {Regula}},\ }\href@noop {} {\bibfield  {journal} {\bibinfo  {journal} {arXiv
  preprint arXiv:1901.08127}\ } (\bibinfo {year} {2019})}\BibitemShut {NoStop}%
\bibitem [{\citenamefont {Napoli}\ \emph {et~al.}(2016)\citenamefont {Napoli},
  \citenamefont {Bromley}, \citenamefont {Cianciaruso}, \citenamefont {Piani},
  \citenamefont {Johnston},\ and\ \citenamefont {Adesso}}]{Napoli2016}%
  \BibitemOpen
  \bibfield  {author} {\bibinfo {author} {\bibfnamefont {C.}~\bibnamefont
  {Napoli}}, \bibinfo {author} {\bibfnamefont {T.~R.}\ \bibnamefont {Bromley}},
  \bibinfo {author} {\bibfnamefont {M.}~\bibnamefont {Cianciaruso}}, \bibinfo
  {author} {\bibfnamefont {M.}~\bibnamefont {Piani}}, \bibinfo {author}
  {\bibfnamefont {N.}~\bibnamefont {Johnston}}, \ and\ \bibinfo {author}
  {\bibfnamefont {G.}~\bibnamefont {Adesso}},\ }\href {\doibase
  10.1103/PhysRevLett.116.150502} {\bibfield  {journal} {\bibinfo  {journal}
  {Phys. Rev. Lett.}\ }\textbf {\bibinfo {volume} {116}},\ \bibinfo {pages}
  {150502} (\bibinfo {year} {2016})}\BibitemShut {NoStop}%
\bibitem [{\citenamefont {Piani}\ and\ \citenamefont
  {Watrous}(2015)}]{Piani2015}%
  \BibitemOpen
  \bibfield  {author} {\bibinfo {author} {\bibfnamefont {M.}~\bibnamefont
  {Piani}}\ and\ \bibinfo {author} {\bibfnamefont {J.}~\bibnamefont
  {Watrous}},\ }\href {\doibase 10.1103/PhysRevLett.114.060404} {\bibfield
  {journal} {\bibinfo  {journal} {Phys. Rev. Lett.}\ }\textbf {\bibinfo
  {volume} {114}},\ \bibinfo {pages} {060404} (\bibinfo {year}
  {2015})}\BibitemShut {NoStop}%
\bibitem [{\citenamefont {Skrzypczyk}\ and\ \citenamefont
  {Linden}(2019)}]{Skrzypczyk2019}%
  \BibitemOpen
  \bibfield  {author} {\bibinfo {author} {\bibfnamefont {P.}~\bibnamefont
  {Skrzypczyk}}\ and\ \bibinfo {author} {\bibfnamefont {N.}~\bibnamefont
  {Linden}},\ }\href {\doibase 10.1103/PhysRevLett.122.140403} {\bibfield
  {journal} {\bibinfo  {journal} {Phys. Rev. Lett.}\ }\textbf {\bibinfo
  {volume} {122}},\ \bibinfo {pages} {140403} (\bibinfo {year}
  {2019})}\BibitemShut {NoStop}%
\bibitem [{\citenamefont {Ducuara}\ and\ \citenamefont
  {Skrzypczyk}(2019{\natexlab{a}})}]{Ducuara2019}%
  \BibitemOpen
  \bibfield  {author} {\bibinfo {author} {\bibfnamefont {A.~F.}\ \bibnamefont
  {Ducuara}}\ and\ \bibinfo {author} {\bibfnamefont {P.}~\bibnamefont
  {Skrzypczyk}},\ }\href@noop {} {\bibfield  {journal} {\bibinfo  {journal}
  {arXiv preprint arXiv:1908.10347}\ } (\bibinfo {year}
  {2019}{\natexlab{a}})}\BibitemShut {NoStop}%
\bibitem [{\citenamefont {Oszmaniec}\ and\ \citenamefont
  {Biswas}(2019)}]{Oszmaniec2019operational}%
  \BibitemOpen
  \bibfield  {author} {\bibinfo {author} {\bibfnamefont {M.}~\bibnamefont
  {Oszmaniec}}\ and\ \bibinfo {author} {\bibfnamefont {T.}~\bibnamefont
  {Biswas}},\ }\href {\doibase 10.22331/q-2019-04-26-133} {\bibfield  {journal}
  {\bibinfo  {journal} {{Quantum}}\ }\textbf {\bibinfo {volume} {3}},\ \bibinfo
  {pages} {133} (\bibinfo {year} {2019})}\BibitemShut {NoStop}%
\bibitem [{\citenamefont {Designolle}\ \emph
  {et~al.}(2019{\natexlab{a}})\citenamefont {Designolle}, \citenamefont
  {Skrzypczyk}, \citenamefont {Fr\"owis},\ and\ \citenamefont
  {Brunner}}]{Designole2018}%
  \BibitemOpen
  \bibfield  {author} {\bibinfo {author} {\bibfnamefont {S.}~\bibnamefont
  {Designolle}}, \bibinfo {author} {\bibfnamefont {P.}~\bibnamefont
  {Skrzypczyk}}, \bibinfo {author} {\bibfnamefont {F.}~\bibnamefont
  {Fr\"owis}}, \ and\ \bibinfo {author} {\bibfnamefont {N.}~\bibnamefont
  {Brunner}},\ }\href {\doibase 10.1103/PhysRevLett.122.050402} {\bibfield
  {journal} {\bibinfo  {journal} {Phys. Rev. Lett.}\ }\textbf {\bibinfo
  {volume} {122}},\ \bibinfo {pages} {050402} (\bibinfo {year}
  {2019}{\natexlab{a}})}\BibitemShut {NoStop}%
\bibitem [{\citenamefont {Designolle}\ \emph
  {et~al.}(2019{\natexlab{b}})\citenamefont {Designolle}, \citenamefont
  {Farkas},\ and\ \citenamefont {Kaniewski}}]{Designolle_2019}%
  \BibitemOpen
  \bibfield  {author} {\bibinfo {author} {\bibfnamefont {S.}~\bibnamefont
  {Designolle}}, \bibinfo {author} {\bibfnamefont {M.}~\bibnamefont {Farkas}},
  \ and\ \bibinfo {author} {\bibfnamefont {J.}~\bibnamefont {Kaniewski}},\
  }\href {\doibase 10.1088/1367-2630/ab5020} {\bibfield  {journal} {\bibinfo
  {journal} {New Journal of Physics}\ }\textbf {\bibinfo {volume} {21}},\
  \bibinfo {pages} {113053} (\bibinfo {year} {2019}{\natexlab{b}})}\BibitemShut
  {NoStop}%
\bibitem [{\citenamefont {Howard}\ and\ \citenamefont
  {Campbell}(2017)}]{Howard2017}%
  \BibitemOpen
  \bibfield  {author} {\bibinfo {author} {\bibfnamefont {M.}~\bibnamefont
  {Howard}}\ and\ \bibinfo {author} {\bibfnamefont {E.}~\bibnamefont
  {Campbell}},\ }\href {\doibase 10.1103/PhysRevLett.118.090501} {\bibfield
  {journal} {\bibinfo  {journal} {Phys. Rev. Lett.}\ }\textbf {\bibinfo
  {volume} {118}},\ \bibinfo {pages} {090501} (\bibinfo {year}
  {2017})}\BibitemShut {NoStop}%
\bibitem [{\citenamefont {Davies}\ and\ \citenamefont
  {Lewis}(1970)}]{Davies1970}%
  \BibitemOpen
  \bibfield  {author} {\bibinfo {author} {\bibfnamefont {E.~B.}\ \bibnamefont
  {Davies}}\ and\ \bibinfo {author} {\bibfnamefont {J.~T.}\ \bibnamefont
  {Lewis}},\ }\href {https://projecteuclid.org:443/euclid.cmp/1103842336}
  {\bibfield  {journal} {\bibinfo  {journal} {Comm. Math. Phys.}\ }\textbf
  {\bibinfo {volume} {17}},\ \bibinfo {pages} {239} (\bibinfo {year}
  {1970})}\BibitemShut {NoStop}%
\bibitem [{\citenamefont {Brand\~ao}\ and\ \citenamefont
  {Gour}(2015)}]{Brandao2015}%
  \BibitemOpen
  \bibfield  {author} {\bibinfo {author} {\bibfnamefont {F.~G. S.~L.}\
  \bibnamefont {Brand\~ao}}\ and\ \bibinfo {author} {\bibfnamefont
  {G.}~\bibnamefont {Gour}},\ }\href {\doibase 10.1103/PhysRevLett.115.070503}
  {\bibfield  {journal} {\bibinfo  {journal} {Phys. Rev. Lett.}\ }\textbf
  {\bibinfo {volume} {115}},\ \bibinfo {pages} {070503} (\bibinfo {year}
  {2015})}\BibitemShut {NoStop}%
\bibitem [{\citenamefont {Bennett}\ \emph {et~al.}(1996)\citenamefont
  {Bennett}, \citenamefont {Bernstein}, \citenamefont {Popescu},\ and\
  \citenamefont {Schumacher}}]{Bennett1996}%
  \BibitemOpen
  \bibfield  {author} {\bibinfo {author} {\bibfnamefont {C.~H.}\ \bibnamefont
  {Bennett}}, \bibinfo {author} {\bibfnamefont {H.~J.}\ \bibnamefont
  {Bernstein}}, \bibinfo {author} {\bibfnamefont {S.}~\bibnamefont {Popescu}},
  \ and\ \bibinfo {author} {\bibfnamefont {B.}~\bibnamefont {Schumacher}},\
  }\href {\doibase 10.1103/PhysRevA.53.2046} {\bibfield  {journal} {\bibinfo
  {journal} {Phys. Rev. A}\ }\textbf {\bibinfo {volume} {53}},\ \bibinfo
  {pages} {2046} (\bibinfo {year} {1996})}\BibitemShut {NoStop}%
\bibitem [{\citenamefont {Gour}\ and\ \citenamefont
  {Spekkens}(2008)}]{Gour2008}%
  \BibitemOpen
  \bibfield  {author} {\bibinfo {author} {\bibfnamefont {G.}~\bibnamefont
  {Gour}}\ and\ \bibinfo {author} {\bibfnamefont {R.~W.}\ \bibnamefont
  {Spekkens}},\ }\href {\doibase 10.1088/1367-2630/10/3/033023} {\bibfield
  {journal} {\bibinfo  {journal} {New Journal of Physics}\ }\textbf {\bibinfo
  {volume} {10}},\ \bibinfo {pages} {033023} (\bibinfo {year}
  {2008})}\BibitemShut {NoStop}%
\bibitem [{\citenamefont {Marvian}\ and\ \citenamefont
  {Spekkens}(2013)}]{Marvian2013}%
  \BibitemOpen
  \bibfield  {author} {\bibinfo {author} {\bibfnamefont {I.}~\bibnamefont
  {Marvian}}\ and\ \bibinfo {author} {\bibfnamefont {R.~W.}\ \bibnamefont
  {Spekkens}},\ }\href {\doibase 10.1088/1367-2630/15/3/033001} {\bibfield
  {journal} {\bibinfo  {journal} {New Journal of Physics}\ }\textbf {\bibinfo
  {volume} {15}},\ \bibinfo {pages} {033001} (\bibinfo {year}
  {2013})}\BibitemShut {NoStop}%
\bibitem [{\citenamefont {Aberg}(2006)}]{Aberg2006}%
  \BibitemOpen
  \bibfield  {author} {\bibinfo {author} {\bibfnamefont {J.}~\bibnamefont
  {Aberg}},\ }\href@noop {} {\bibfield  {journal} {\bibinfo  {journal} {arXiv
  preprint quant-ph/0612146}\ } (\bibinfo {year} {2006})}\BibitemShut {NoStop}%
\bibitem [{\citenamefont {Baumgratz}\ \emph {et~al.}(2014)\citenamefont
  {Baumgratz}, \citenamefont {Cramer},\ and\ \citenamefont
  {Plenio}}]{Baumgratz2014}%
  \BibitemOpen
  \bibfield  {author} {\bibinfo {author} {\bibfnamefont {T.}~\bibnamefont
  {Baumgratz}}, \bibinfo {author} {\bibfnamefont {M.}~\bibnamefont {Cramer}}, \
  and\ \bibinfo {author} {\bibfnamefont {M.~B.}\ \bibnamefont {Plenio}},\
  }\href {\doibase 10.1103/PhysRevLett.113.140401} {\bibfield  {journal}
  {\bibinfo  {journal} {Phys. Rev. Lett.}\ }\textbf {\bibinfo {volume} {113}},\
  \bibinfo {pages} {140401} (\bibinfo {year} {2014})}\BibitemShut {NoStop}%
\bibitem [{\citenamefont {Horodecki}\ \emph
  {et~al.}(2003{\natexlab{b}})\citenamefont {Horodecki}, \citenamefont
  {Horodecki},\ and\ \citenamefont {Oppenheim}}]{Horodecki2003a}%
  \BibitemOpen
  \bibfield  {author} {\bibinfo {author} {\bibfnamefont {M.}~\bibnamefont
  {Horodecki}}, \bibinfo {author} {\bibfnamefont {P.}~\bibnamefont
  {Horodecki}}, \ and\ \bibinfo {author} {\bibfnamefont {J.}~\bibnamefont
  {Oppenheim}},\ }\href {\doibase 10.1103/PhysRevA.67.062104} {\bibfield
  {journal} {\bibinfo  {journal} {Phys. Rev. A}\ }\textbf {\bibinfo {volume}
  {67}},\ \bibinfo {pages} {062104} (\bibinfo {year}
  {2003}{\natexlab{b}})}\BibitemShut {NoStop}%
\bibitem [{\citenamefont {Janzing}\ \emph {et~al.}(2000)\citenamefont
  {Janzing}, \citenamefont {Wocjan}, \citenamefont {Zeier}, \citenamefont
  {Geiss},\ and\ \citenamefont {Beth}}]{Janzing2000}%
  \BibitemOpen
  \bibfield  {author} {\bibinfo {author} {\bibfnamefont {D.}~\bibnamefont
  {Janzing}}, \bibinfo {author} {\bibfnamefont {P.}~\bibnamefont {Wocjan}},
  \bibinfo {author} {\bibfnamefont {R.}~\bibnamefont {Zeier}}, \bibinfo
  {author} {\bibfnamefont {R.}~\bibnamefont {Geiss}}, \ and\ \bibinfo {author}
  {\bibfnamefont {T.}~\bibnamefont {Beth}},\ }\href {\doibase
  10.1023/A:1026422630734} {\bibfield  {journal} {\bibinfo  {journal}
  {International Journal of Theoretical Physics}\ }\textbf {\bibinfo {volume}
  {39}},\ \bibinfo {pages} {2717} (\bibinfo {year} {2000})}\BibitemShut
  {NoStop}%
\bibitem [{\citenamefont {Horodecki}\ and\ \citenamefont
  {Oppenheim}(2013)}]{Horodecki2013}%
  \BibitemOpen
  \bibfield  {author} {\bibinfo {author} {\bibfnamefont {M.}~\bibnamefont
  {Horodecki}}\ and\ \bibinfo {author} {\bibfnamefont {J.}~\bibnamefont
  {Oppenheim}},\ }\href {https://doi.org/10.1038/ncomms3059} {\bibfield
  {journal} {\bibinfo  {journal} {Nat. Commun.}\ }\textbf {\bibinfo {volume}
  {4}},\ \bibinfo {pages} {2059} (\bibinfo {year} {2013})}\BibitemShut
  {NoStop}%
\bibitem [{\citenamefont {Brand\~ao}\ \emph {et~al.}(2013)\citenamefont
  {Brand\~ao}, \citenamefont {Horodecki}, \citenamefont {Oppenheim},
  \citenamefont {Renes},\ and\ \citenamefont {Spekkens}}]{Brandao2013}%
  \BibitemOpen
  \bibfield  {author} {\bibinfo {author} {\bibfnamefont {F.~G. S.~L.}\
  \bibnamefont {Brand\~ao}}, \bibinfo {author} {\bibfnamefont {M.}~\bibnamefont
  {Horodecki}}, \bibinfo {author} {\bibfnamefont {J.}~\bibnamefont
  {Oppenheim}}, \bibinfo {author} {\bibfnamefont {J.~M.}\ \bibnamefont
  {Renes}}, \ and\ \bibinfo {author} {\bibfnamefont {R.~W.}\ \bibnamefont
  {Spekkens}},\ }\href {\doibase 10.1103/PhysRevLett.111.250404} {\bibfield
  {journal} {\bibinfo  {journal} {Phys. Rev. Lett.}\ }\textbf {\bibinfo
  {volume} {111}},\ \bibinfo {pages} {250404} (\bibinfo {year}
  {2013})}\BibitemShut {NoStop}%
\bibitem [{\citenamefont {Veitch}\ \emph {et~al.}(2014)\citenamefont {Veitch},
  \citenamefont {Mousavian}, \citenamefont {Gottesman},\ and\ \citenamefont
  {Emerson}}]{Veitch2014}%
  \BibitemOpen
  \bibfield  {author} {\bibinfo {author} {\bibfnamefont {V.}~\bibnamefont
  {Veitch}}, \bibinfo {author} {\bibfnamefont {S.~A.~H.}\ \bibnamefont
  {Mousavian}}, \bibinfo {author} {\bibfnamefont {D.}~\bibnamefont
  {Gottesman}}, \ and\ \bibinfo {author} {\bibfnamefont {J.}~\bibnamefont
  {Emerson}},\ }\href {\doibase 10.1088/1367-2630/16/1/013009} {\bibfield
  {journal} {\bibinfo  {journal} {New Journa of Physics}\ }\textbf {\bibinfo
  {volume} {16}},\ \bibinfo {pages} {013009} (\bibinfo {year}
  {2014})}\BibitemShut {NoStop}%
\bibitem [{\citenamefont {de~Vicente}(2014)}]{deVicente2014}%
  \BibitemOpen
  \bibfield  {author} {\bibinfo {author} {\bibfnamefont {J.~I.}\ \bibnamefont
  {de~Vicente}},\ }\href {\doibase 10.1088/1751-8113/47/42/424017} {\bibfield
  {journal} {\bibinfo  {journal} {Journal of Physics A: Mathematical and
  Theoretical}\ }\textbf {\bibinfo {volume} {47}},\ \bibinfo {pages} {424017}
  (\bibinfo {year} {2014})}\BibitemShut {NoStop}%
\bibitem [{\citenamefont {Gallego}\ and\ \citenamefont
  {Aolita}(2015)}]{Gallego2015}%
  \BibitemOpen
  \bibfield  {author} {\bibinfo {author} {\bibfnamefont {R.}~\bibnamefont
  {Gallego}}\ and\ \bibinfo {author} {\bibfnamefont {L.}~\bibnamefont
  {Aolita}},\ }\href {\doibase 10.1103/PhysRevX.5.041008} {\bibfield  {journal}
  {\bibinfo  {journal} {Phys. Rev. X}\ }\textbf {\bibinfo {volume} {5}},\
  \bibinfo {pages} {041008} (\bibinfo {year} {2015})}\BibitemShut {NoStop}%
\bibitem [{\citenamefont {Horodecki}\ \emph {et~al.}(2015)\citenamefont
  {Horodecki}, \citenamefont {Grudka}, \citenamefont {Joshi}, \citenamefont
  {K\l{}obus},\ and\ \citenamefont {\L{}odyga}}]{Horodecki2015}%
  \BibitemOpen
  \bibfield  {author} {\bibinfo {author} {\bibfnamefont {K.}~\bibnamefont
  {Horodecki}}, \bibinfo {author} {\bibfnamefont {A.}~\bibnamefont {Grudka}},
  \bibinfo {author} {\bibfnamefont {P.}~\bibnamefont {Joshi}}, \bibinfo
  {author} {\bibfnamefont {W.}~\bibnamefont {K\l{}obus}}, \ and\ \bibinfo
  {author} {\bibfnamefont {J.}~\bibnamefont {\L{}odyga}},\ }\href {\doibase
  10.1103/PhysRevA.92.032104} {\bibfield  {journal} {\bibinfo  {journal} {Phys.
  Rev. A}\ }\textbf {\bibinfo {volume} {92}},\ \bibinfo {pages} {032104}
  (\bibinfo {year} {2015})}\BibitemShut {NoStop}%
\bibitem [{\citenamefont {Chitambar}\ and\ \citenamefont
  {Gour}(2019)}]{Chitambar2019}%
  \BibitemOpen
  \bibfield  {author} {\bibinfo {author} {\bibfnamefont {E.}~\bibnamefont
  {Chitambar}}\ and\ \bibinfo {author} {\bibfnamefont {G.}~\bibnamefont
  {Gour}},\ }\href {\doibase 10.1103/RevModPhys.91.025001} {\bibfield
  {journal} {\bibinfo  {journal} {Rev. Mod. Phys.}\ }\textbf {\bibinfo {volume}
  {91}},\ \bibinfo {pages} {025001} (\bibinfo {year} {2019})}\BibitemShut
  {NoStop}%
\bibitem [{\citenamefont {Liu}\ \emph {et~al.}(2019)\citenamefont {Liu},
  \citenamefont {Bu},\ and\ \citenamefont {Takagi}}]{Liu2019}%
  \BibitemOpen
  \bibfield  {author} {\bibinfo {author} {\bibfnamefont {Z.-W.}\ \bibnamefont
  {Liu}}, \bibinfo {author} {\bibfnamefont {K.}~\bibnamefont {Bu}}, \ and\
  \bibinfo {author} {\bibfnamefont {R.}~\bibnamefont {Takagi}},\ }\href
  {\doibase 10.1103/PhysRevLett.123.020401} {\bibfield  {journal} {\bibinfo
  {journal} {Phys. Rev. Lett.}\ }\textbf {\bibinfo {volume} {123}},\ \bibinfo
  {pages} {020401} (\bibinfo {year} {2019})}\BibitemShut {NoStop}%
\bibitem [{\citenamefont {Liu}\ and\ \citenamefont {Yuan}(2019)}]{Liu2019op}%
  \BibitemOpen
  \bibfield  {author} {\bibinfo {author} {\bibfnamefont {Y.}~\bibnamefont
  {Liu}}\ and\ \bibinfo {author} {\bibfnamefont {X.}~\bibnamefont {Yuan}},\
  }\href@noop {} {\bibfield  {journal} {\bibinfo  {journal} {arXiv preprint
  arXiv:1904.02680}\ } (\bibinfo {year} {2019})}\BibitemShut {NoStop}%
\bibitem [{\citenamefont {Theurer}\ \emph {et~al.}(2019)\citenamefont
  {Theurer}, \citenamefont {Egloff}, \citenamefont {Zhang},\ and\ \citenamefont
  {Plenio}}]{Theurer2019}%
  \BibitemOpen
  \bibfield  {author} {\bibinfo {author} {\bibfnamefont {T.}~\bibnamefont
  {Theurer}}, \bibinfo {author} {\bibfnamefont {D.}~\bibnamefont {Egloff}},
  \bibinfo {author} {\bibfnamefont {L.}~\bibnamefont {Zhang}}, \ and\ \bibinfo
  {author} {\bibfnamefont {M.~B.}\ \bibnamefont {Plenio}},\ }\href {\doibase
  10.1103/PhysRevLett.122.190405} {\bibfield  {journal} {\bibinfo  {journal}
  {Phys. Rev. Lett.}\ }\textbf {\bibinfo {volume} {122}},\ \bibinfo {pages}
  {190405} (\bibinfo {year} {2019})}\BibitemShut {NoStop}%
\bibitem [{\citenamefont {Uola}\ \emph {et~al.}(2019)\citenamefont {Uola},
  \citenamefont {Kraft},\ and\ \citenamefont {Abbott}}]{Uola2019}%
  \BibitemOpen
  \bibfield  {author} {\bibinfo {author} {\bibfnamefont {R.}~\bibnamefont
  {Uola}}, \bibinfo {author} {\bibfnamefont {T.}~\bibnamefont {Kraft}}, \ and\
  \bibinfo {author} {\bibfnamefont {A.~A.}\ \bibnamefont {Abbott}},\
  }\href@noop {} {\bibfield  {journal} {\bibinfo  {journal} {arXiv preprint
  arXiv:1906.09206}\ } (\bibinfo {year} {2019})}\BibitemShut {NoStop}%
\bibitem [{\citenamefont {Ducuara}\ and\ \citenamefont
  {Skrzypczyk}(2019{\natexlab{b}})}]{Ducuara2019a}%
  \BibitemOpen
  \bibfield  {author} {\bibinfo {author} {\bibfnamefont {A.~F.}\ \bibnamefont
  {Ducuara}}\ and\ \bibinfo {author} {\bibfnamefont {P.}~\bibnamefont
  {Skrzypczyk}},\ }\href@noop {} {\bibfield  {journal} {\bibinfo  {journal}
  {arXiv preprint arXiv:1909.10486}\ } (\bibinfo {year}
  {2019}{\natexlab{b}})}\BibitemShut {NoStop}%
\bibitem [{\citenamefont {Boyd}\ and\ \citenamefont
  {Vandenberghe}(2004)}]{Boyd2004}%
  \BibitemOpen
  \bibfield  {author} {\bibinfo {author} {\bibfnamefont {S.}~\bibnamefont
  {Boyd}}\ and\ \bibinfo {author} {\bibfnamefont {L.}~\bibnamefont
  {Vandenberghe}},\ }\href@noop {} {\emph {\bibinfo {title} {Convex
  Optimization}}}\ (\bibinfo  {publisher} {Cambridge University Press},\
  \bibinfo {address} {New York, NY, USA},\ \bibinfo {year} {2004})\BibitemShut
  {NoStop}%
\bibitem [{\citenamefont {Peres}(1996)}]{Peres94}%
  \BibitemOpen
  \bibfield  {author} {\bibinfo {author} {\bibfnamefont {A.}~\bibnamefont
  {Peres}},\ }\href {\doibase 10.1103/PhysRevLett.77.1413} {\bibfield
  {journal} {\bibinfo  {journal} {Phys. Rev. Lett.}\ }\textbf {\bibinfo
  {volume} {77}},\ \bibinfo {pages} {1413} (\bibinfo {year}
  {1996})}\BibitemShut {NoStop}%
\bibitem [{\citenamefont {Kraus}(1971)}]{Kraus1971}%
  \BibitemOpen
  \bibfield  {author} {\bibinfo {author} {\bibfnamefont {K.}~\bibnamefont
  {Kraus}},\ }\href {\doibase https://doi.org/10.1016/0003-4916(71)90108-4}
  {\bibfield  {journal} {\bibinfo  {journal} {Annals of Physics}\ }\textbf
  {\bibinfo {volume} {64}},\ \bibinfo {pages} {311 } (\bibinfo {year}
  {1971})}\BibitemShut {NoStop}%
\bibitem [{\citenamefont {Jamiołkowski}(1972)}]{Jamiolkowski1972}%
  \BibitemOpen
  \bibfield  {author} {\bibinfo {author} {\bibfnamefont {A.}~\bibnamefont
  {Jamiołkowski}},\ }\href {\doibase
  https://doi.org/10.1016/0034-4877(72)90011-0} {\bibfield  {journal} {\bibinfo
   {journal} {Reports on Mathematical Physics}\ }\textbf {\bibinfo {volume}
  {3}},\ \bibinfo {pages} {275 } (\bibinfo {year} {1972})}\BibitemShut
  {NoStop}%
\bibitem [{\citenamefont {Choi}(1975)}]{Choi1975}%
  \BibitemOpen
  \bibfield  {author} {\bibinfo {author} {\bibfnamefont {M.-D.}\ \bibnamefont
  {Choi}},\ }\href {\doibase https://doi.org/10.1016/0024-3795(75)90075-0}
  {\bibfield  {journal} {\bibinfo  {journal} {Linear Algebra and its
  Applications}\ }\textbf {\bibinfo {volume} {10}},\ \bibinfo {pages} {285 }
  (\bibinfo {year} {1975})}\BibitemShut {NoStop}%
\bibitem [{\citenamefont {Wilde}(2013)}]{Wilde2013}%
  \BibitemOpen
  \bibfield  {author} {\bibinfo {author} {\bibfnamefont {M.~M.}\ \bibnamefont
  {Wilde}},\ }\href {\doibase 10.1017/CBO9781139525343} {\emph {\bibinfo
  {title} {Quantum Information Theory}}}\ (\bibinfo  {publisher} {Cambridge
  University Press},\ \bibinfo {year} {2013})\BibitemShut {NoStop}%
\bibitem [{\citenamefont {Horodecki}\ \emph {et~al.}(2009)\citenamefont
  {Horodecki}, \citenamefont {Horodecki}, \citenamefont {Horodecki},\ and\
  \citenamefont {Horodecki}}]{Horodecki2009}%
  \BibitemOpen
  \bibfield  {author} {\bibinfo {author} {\bibfnamefont {R.}~\bibnamefont
  {Horodecki}}, \bibinfo {author} {\bibfnamefont {P.}~\bibnamefont
  {Horodecki}}, \bibinfo {author} {\bibfnamefont {M.}~\bibnamefont
  {Horodecki}}, \ and\ \bibinfo {author} {\bibfnamefont {K.}~\bibnamefont
  {Horodecki}},\ }\href {\doibase 10.1103/RevModPhys.81.865} {\bibfield
  {journal} {\bibinfo  {journal} {Rev. Mod. Phys.}\ }\textbf {\bibinfo {volume}
  {81}},\ \bibinfo {pages} {865} (\bibinfo {year} {2009})}\BibitemShut
  {NoStop}%
\bibitem [{\citenamefont {Coecke}\ and\ \citenamefont
  {Kissinger}(2017)}]{Coecke2017}%
  \BibitemOpen
  \bibfield  {author} {\bibinfo {author} {\bibfnamefont {B.}~\bibnamefont
  {Coecke}}\ and\ \bibinfo {author} {\bibfnamefont {A.}~\bibnamefont
  {Kissinger}},\ }\href@noop {} {\emph {\bibinfo {title} {Picturing quantum
  processes}}}\ (\bibinfo  {publisher} {Cambridge University Press},\ \bibinfo
  {year} {2017})\BibitemShut {NoStop}%
\bibitem [{\citenamefont {Wood}\ \emph {et~al.}(2011)\citenamefont {Wood},
  \citenamefont {Biamonte},\ and\ \citenamefont {Cory}}]{Wood2011}%
  \BibitemOpen
  \bibfield  {author} {\bibinfo {author} {\bibfnamefont {C.~J.}\ \bibnamefont
  {Wood}}, \bibinfo {author} {\bibfnamefont {J.~D.}\ \bibnamefont {Biamonte}},
  \ and\ \bibinfo {author} {\bibfnamefont {D.~G.}\ \bibnamefont {Cory}},\
  }\href@noop {} {\bibfield  {journal} {\bibinfo  {journal} {arXiv preprint
  arXiv:1111.6950}\ } (\bibinfo {year} {2011})}\BibitemShut {NoStop}%
\bibitem [{\citenamefont {Biamonte}\ and\ \citenamefont
  {Bergholm}(2017)}]{Biamonte2017}%
  \BibitemOpen
  \bibfield  {author} {\bibinfo {author} {\bibfnamefont {J.}~\bibnamefont
  {Biamonte}}\ and\ \bibinfo {author} {\bibfnamefont {V.}~\bibnamefont
  {Bergholm}},\ }\href@noop {} {\bibfield  {journal} {\bibinfo  {journal}
  {arXiv preprint arXiv:1708.00006}\ } (\bibinfo {year} {2017})}\BibitemShut
  {NoStop}%
\end{thebibliography}%

\onecolumngrid
\appendix

\section{Notation and useful facts}
In what follows we restrict all quantum systems (A, B, etc.) to be associated with finite-dimensional Hilbert spaces ($\mathcal{H}_{\text{A}}$, $\mathcal{H}_{\text{B}}$, etc.). The maximally-entangled state for two orthonormal basis sets $\{\ket{i}_{\text{A}}\}$ and $\{\ket{i}_{\text{B}}\}$ for $\mathcal{H}_{\text{A}}$ and $\mathcal{H}_{\text{B}}$ with $\dim \mathcal{H}_{\text{A}}=\dim \mathcal{H}_{\text{B}} = d$ respectively, will be denoted as:
\begin{align}
    \ket{\phi_+}_{\text{AB}} = \frac{1}{\sqrt{d}} \sum_{i=1}^d \ket{i}_{\text{A}} \ot \ket{i}_{\text{B}}.
\end{align}
Transpose map $\mathcal{T}$ acting between linear operators is defined as:
\begin{align}
    \mathcal{T}(X) = X^{T},
\end{align}
where $X^T$ denotes the ordinary transpose of matrix $X$. Partial transpose of a bipartite operator $X^{\text{AB}}$ with respect to subsystem $A$ is denoted $(X^{\text{AB}})^{T_{\text{A}}}$ and defined as:
\begin{align}
    (X^{\text{AB}})^{T_{\text{A}}}:= (\mathcal{T}^{\text{A}} \ot \mathcal{I}^{\text{B}}) [X^{\text{AB}}],
\end{align}
and similarly for subsystem B. If a state $\rho$ is separable, denoted $\rho \in \sep$, then its density operator has a positive partial transpose. In that case we call it a PPT state \cite{Peres94}, or in the case of general operators --- a PPT operator. Transpose of a linear map $\mathcal{E}$ is denoted $\mathcal{E}^{T}$ and defined as:
\begin{align}
    \mathcal{E}^T[X] := \mathcal{T} \circ \mathcal{E}\, [X],
\end{align}
where $\circ$ denotes composition of maps. The adjoint of $\mathcal{E}$ is defined to be the unique map $\mathcal{E}^{\dagger}$ which satisfies:
\begin{align}
    \tr\left[X \cdot \mathcal{E}^{\dagger}[Y]\right] = \tr\left[\mathcal{E}[X]\cdot Y\right]
\end{align}
In what follows we will make extensive use of several important properties of the maximally-entangled state $\ket{\phi_+}_{\text{AB}}$. The first of them holds for an arbitrary linear operator $E$:
\begin{align}
    \label{eq:swapMEV}
    (\mathbb{1}^{\text{A}} \ot E^{\text{B}}) \ket{\phi_+}_{\text{AB}} = ((E^{\text{A}})^{T} \ot \mathbb{1}^{\text{B}}) \ket{\phi_+}_{\text{AB}}
\end{align}
Using Kraus decomposition of a linear map \cite{Kraus1971} it can be further shown that the following holds for an arbitrary linear map $\mathcal{E}$:
\begin{align}
    \label{eq_prop_ment}
    (\mathcal{I}^{\text{A}} \ot \mathcal{E}^{\text{B}})\, \phi_{+}^{\text{AB}} = ((\mathcal{E}^{\text{A}})^{T} \ot \mathcal{I}^{\text{B}})\, [\phi_+^{\text{AB}}],
\end{align}
where we denoted $\phi_+^{\text{AB}} := \dyad{\phi_+}_{\text{AB}}$. Another identity which will be utilized frequently in this Appendix is given by:
\begin{align}
    \label{eq:tran_id}
    \tr_{\text{B}} [(\phi_+^{\text{AB}} \ot \mathbb{1}^{\text{C}}) (\mathbb{1}^{\text{A}} \ot X^{\text{BC}})] = \frac{1}{d}\, (X^{\text{AC}})^{T_{\text{A}}},
\end{align}
which holds for an arbitrary bipartite linear operator $X$. Similarly, we will also use the identity:
\begin{align}
    \label{snake_id}
    \tr_{\text{CD}}\left[(\mathbb{1}^{\text{A}} \ot \phi_+^{\text{CD}} \ot \mathbb{1}^{\text{D}}) (X^{\text{AC}} \ot \phi_+^{\text{D}\text{B}}) \right] = \frac{1}{d^2}\,X^{\text{AB}},
\end{align} 
which is again valid for an arbitrary bipartite linear operator $X$. Finally, the Choi-Jamiołkowski operator of a linear map $\mathcal{E}$ is given by:
\begin{align}   
    \label{eq:cj_mat}
    J_{\mathcal{E}} := (\mathcal{I}^{\text{A}} \ot \mathcal{E}^{\text{B}}) \, [\phi_+^{\text{AB}}]
\end{align}
The map $\mathcal{E}$ is completely-positive if and only if $J_{\mathcal{E}} \geq 0$ and trace-preserving if and only if $\tr_{\text{B}}[J_{\mathcal{E}}] = \mathbb{1}^{\text{A}}$. The action of map $\mathcal{E}$ on operator $X$ is fully specified by the Choi-Jamiołkowski operator and given by:
\begin{align}
    \label{eq:cj_chann}
    \mathcal{E}[X] = \tr_{\text{A}}\left[ \left((X^{\text{A}})^T \ot \mathbb{1}^{\text{B}}\right) J_{\mathcal{E}}^{\text{AB}} \right]
\end{align}

\section{Equivalent formulation for the Robustness of Teleportation}
Let us start with the definition of the optimization problem (\ref{eq:5}). Our first goal is to rewrite the first line of constraints in (\ref{eq:5}) using  Choi-Jamiołkowski operators: $J_a := (\mathcal{I} \ot \Lambda_a)[\phi_+]$, $R_a := (\mathcal{I} \ot \Lambda_a')[\phi_+]$ and $F_a := (\mathcal{I} \ot \Lambda_a^{c})[\phi_+]$. Notice that using the Choi-Jamiołkowski isomorphism \cite{Jamiolkowski1972,Choi1975} (see (\ref{eq:cj_chann})) the first line of constraints in (\ref{eq:5}) can be equivalently written as:
\begin{align}
    \label{eq:chann_to_choi}
    \forall \, a, x \qquad \tr_{\text{V}}\left[\left((\omega_x^{\text{V}})^T \ot \mathbb{1}^{\text{B}} \right)\left(\frac{1}{1+r} J_a^{\text{VB}} + \frac{r}{1+r} R_a^{\text{VB}}\right) \right] = \tr_{\text{V}}\left[\left((\omega_x^{\text{V}})^T \ot \mathbb{1}^{\text{B}} \right)\left(F_a^{\text{VB}}\right) \right] 
\end{align}
Let us now assume that the set of input states $\{\omega_x\}$ form a tomographically-complete set. This means that any density matrix $\omega'$ can be expressed as a linear combination of states from the set $\{\omega_x\}$, i.e. $\omega' = \sum_{x} p(x)\, \omega_x$ for some probability distribution $p(x)$. In this case (\ref{eq:chann_to_choi}) can only be satisfied if:
\begin{align}
    \forall \, a \qquad \frac{1}{1+r} J_a^{\text{VB}} + \frac{r}{1+r} R_a^{\text{VB}} = F_a^{\text{VB}}
\end{align}
The original problem (\ref{eq:5}) can be then equivalently rewritten as:
\begin{align} 
    \label{eq:99}
\mathcal{T}(\mathbb{\Lambda}, \{\omega_x\}_x) = \qquad \min_{\{{R}_a\}, \{{F}_a\}, r} \quad & r \\ \nonumber
     & \frac{1}{1+r} J_a + \frac{r}{1+r} {R}_a = {F}_a  \qquad \qquad \qquad \quad \forall\, a \\ \nonumber
    &  \{\Lambda^c_a\} \in \mathcal{F}, \quad \{\Lambda'_a\} \in \mathcal{R} \qquad \qquad \qquad \qquad \forall \, a \nonumber
\end{align}
Notice that in this form the optimization problem (\ref{eq:5}) becomes effectively independent of $\{\omega_x\}$ which we will denote by writing $\mathcal{T}(\Lam{})$ instead of $\mathcal{T}(\mathbb{\Lambda}, \{\omega_x\}_x)$.  Our next goal is to write the constraints on the instruments $\{\Lambda_a'\}$ and $\{\Lambda_a^c\}$ in  (\ref{eq:99}) in terms of equivalent constraints on the associated Choi-Jamiołkowski operators $\{R_a\}$ and $\{F_a\}$.

Before we proceed let us recall the general form of the teleportation instrument $\Lam{} = \{\Lambda_a\}$ which is formed from a measurement $\{M_a\}$ and a shared state $\rho^{\text{AB}}$:
\begin{align}
    \Lambda_a^{\text{V}'\rightarrow\text{B}}[\omega] := \tr_{\text{V}'\text{A}}\left[(M_a^{\text{V}'\text{A}} \ot \mathbb{1}^{\text{B}} )(\omega^{\text{V}'} \ot \rho^{\text{AB}})\right]
\end{align}
The associated Choi-Jamiołkowski operators are given by:
\begin{align}
    J_a^{\text{V}\text{B}} = (\mathcal{I}^{\text{V}} \ot \Lambda_a^{\text{V}'\rightarrow\text{B}})[\phi_+^{\text{V}\text{V}'}] = \tr_{\text{V}'\text{A}}\left[(\mathbb{1}^{\text{V}} \ot M_a^{\text{V}'\text{A}} \ot \mathbb{1}^{\text{B}}) (\phi_{+}^{\text{V}\text{V}'} \ot \rho^{\text{AB}})\right].
\end{align}
Let us now characterize these operators for the case of arbitrary and classical teleportation instruments.
\paragraph{Characterization of $\{R_a\}$.} 
The constraint $\{\Lambda_a'\} \in \mathcal{R}$ means that $\{\Lambda_a'\}$ forms an arbitrary teleportation instrument. Let $\{R_a^{\text{V}\text{B}}\}$ be the set of the associated Choi-Jamiołkowski operators, i.e.:
\begin{align}
    \label{eq:jap}
    R_a^{\text{V}\text{B}} = \tr_{\text{V}'\text{A}}\left[(\mathbb{1}^{\text{V}} \ot N_a^{\text{V}'\text{A}} \ot \mathbb{1}^{\text{B}}) (\phi_{+}^{\text{V}\text{V}'} \ot \eta^{\text{AB}})\right],
\end{align}
where $\{N_a^{\text{V}'\text{A}}\}$ can be any bipartite POVM and $\eta^{\text{AB}}$ can be any bipartite state. By inspection we can see that $\{R_a^{\text{VB}}\}$ are $(i)$ positive for all $a$ and $(ii)$ satisfy the no-signalling condition $\sum_a R_a^{\text{V}\text{B}} = {d_{\text{V}}}^{-1} \cdot \mathbb{1}^{\text{V}} \ot \eta^{\text{B}}$. These are not only necessary, but also sufficient conditions for a set of operators to be Choi-Jamiołkowski operators of some teleportation instrument. In other words, any family of operators satisfying $(i)$ and $(ii)$ can be written in the form (\ref{eq:jap}). 

To see this, consider an arbitrary set of positive operators $\{X_a\}$ satisfying  $\sum_a X_a^{\text{V}\text{B}} = {d_{\text{V}}}^{-1}\cdot \mathbb{1}^{\text{V}}\ot \eta^{\text{B}}$ and let $\ket{\eta}_{\text{AB}}$ be the purification \cite{Wilde2013} of $\eta^{\text{B}}$, i.e.:
\begin{align}
    \ket{\eta}_{\text{AB}} = \sqrt{d_{\text{A}}}\, (\mathbb{1}^{\text{A}}\ot {\eta}_{\text{\,B}}^{1/2}) \ket{\phi_+}_{\text{AB}}.
\end{align}
We now make the following choice of operators in (\ref{eq:jap}):
\begin{align}
    \eta^{\text{AB}} = \dyad{\eta}_{\text{AB}} \qquad  N_a^{\text{V}'\text{A}} = d_{\text{V}'} \, \left[ \mathbb{1}^{\text{V}'} \ot (\eta^{-1/2}_{\text{A}})^{T}  \right] (X_a^{\text{V}'\text{A}})^T \left[ \mathbb{1}^{\text{V}'} \ot (\eta^{-1/2}_{\text{A}})^{T}  \right],
\end{align}
It can be easily verified that the operators $\{N_a^{\text{VA}}\}$ form a POVM. Moreover, by plugging these into (\ref{eq:jap}) we obtain:
\begin{align}
    \label{eq:Ra1}
   R_a^{\text{VB}} &= d_{\text{A}} \, d_{\text{V}'}\cdot \tr_{\text{V}'\text{A}}\!\! \left[\left(\mathbb{1}^{\text{V}} \! \ot \! 
    \left( \mathbb{1}^{\text{V}'} \!\! \ot  (\eta^{-1/2}_{\text{A}})^{T}  \right) \! \! \left(X_a^{\text{V}'\text{A}}\right)^{\!T}\!\!\! \left( \mathbb{1}^{\text{V}'}\! \ot (\eta^{-1/2}_{\text{A}})^{T}  \right) \!
   \ot \mathbb{1}^{\text{B}}\!\! \right)\!\! \left(\phi_{+}^{\text{V}\text{V}'}\!\! \ot (\mathbb{1}^{\text{A}} \ot \eta_{\text{B}}^{1/2}) \phi_+^{\text{AB}} (\mathbb{1}^{\text{A}} \ot \eta_{\text{B}}^{1/2}) \right)\right]\\
   &= d_{\text{A}} \, d_{\text{V}'}\cdot   \tr_{\text{V}'\text{A}}\left[(\mathbb{1}^{\text{V}} \ot (X_a^{\text{V}'\text{A}})^T \ot \mathbb{1}^{\text{B}}) (\phi_{+}^{\text{V}\text{V}'} \ot \phi_+^{\text{AB}})\right] \\
    &= X_a^{\text{V}\text{B}} \label{eq:Ra2}
\end{align}
where in the first line we used (\ref{eq:swapMEV}) and in the second one we applied twice (\ref{eq:tran_id}). Since $X_a^{\text{V}\text{B}}$ was by assumption an arbitrary set of operators satisfying $(i)$ and $(ii)$ we conclude that any such operator can be written as $R_a^{\text{V}\text{B}}$ for some choice of measurement and shared state. Hence we have the following equivalence: 
\begin{align}
    \label{eq:equivR}
    \{\Lambda_a'\} \in \mathcal{R} \iff R_a^{\text{V}\text{B}} \geq 0 \quad \text{and} \quad \sum_a R_a^{\text{V}\text{B}} =  \frac{1}{d_{\text{V}}}\, \mathbb{1}^{\text{V}} \ot \eta^{\text{B}}.
\end{align}

\paragraph{Characterization of $\{F_a\}$.} 
The constraint $\{\Lambda_a^{c}\} \in \mathcal{F}$ means that the teleportation instrument $\Lam{}^c = \{\Lambda_a^{c}\}$ is classical, i.e. it arises either from a separable shared state or a separable measurement. Let us denote the state with $\tau^{\text{AB}}$ and the measurement with $O_a^{\text{VA}}$. The associated Choi-Jamiołkowski operator has the following form:
\begin{align}
    \label{eq:fa}
    F_a^{\text{V}\text{B}} = \tr_{\text{V}'\text{A}}\left[(\mathbb{1}^{\text{V}} \ot O_a^{\text{V}'\text{A}} \ot \mathbb{1}^{\text{B}}) (\phi_{+}^{\text{V}\text{V}'} \ot \tau^{\text{AB}})\right],
\end{align}
where either $\tau^{\text{AB}} \in \sep$ or $O_a^{\text{V}'\text{A}} \in \sep \, \forall \, a$. Since the analysis of these two cases is essentially the same we will just consider the case when $\tau^{\text{AB}}$ is separable. The most general separable state $\tau^{\text{AB}}$ can be written as:
\begin{align}
    \tau^{\text{AB}} = \sum_{\lambda} p_{\lambda}\, \tau_{\lambda}^{\text{A}} \ot \tau_{\lambda}^{\text{B}},
\end{align} 
This means that operators $F_a^{\text{VB}}$ take the form:
\begin{align}
    \label{eq:Fasep}
    F_a^{\text{VB}} &= \sum_{\lambda} p_{\lambda} \tr_{\text{V}'\text{A}} [(\mathbb{1}^{\text{V}} \ot O_a^{\text{V}'\text{A}} \ot \mathbb{1}^{\text{B}})  (\phi_+^{\text{V}\text{V}'} \ot \tau_{\lambda}^{\text{A}} \ot \tau_{\lambda}^{\text{B}})] \\
    &= \sum_{\lambda} p_{\lambda}\, O_{a|\lambda}^{\text{V}} \ot \tau_{\lambda}^{\text{B}},
\end{align}
where $O_{a|\lambda}^{\text{V}} = \tr_{\text{V}'\text{A}} [(\mathbb{1}^{\text{V}} \ot O_a^{\text{V}'\text{A}}) (\phi_+^{\text{V}\text{V}'} \ot \tau_{\lambda}^{\text{A}})]$. This implies that the operators $F_a^{\text{VB}}$ ($i$) sum up to ${d_{\text{V}}^{-1}} \cdot \mathbb{1}^{\text{A}} \ot \tau^{\text{B}}$ where $\tau^{\text{B}} = \tr_{\text{A}}[\tau^{\text{AB}}]$ and ($ii$) are separable operators. Similarly as before we now infer that any family of operators satisfying $(i)$ and $(ii)$ can be written as in (\ref{eq:Fasep}). Let us assume that $\{Y_a^{\text{VB}}\}$ is such a family. Following similar steps as in the case $(a)$ we take $\ket{\tau}_{\text{AB}}$ to be the purification of $\tau^{\text{B}}$, i.e.:
\begin{align}
    \ket{\tau}_{\text{AB}} = \sqrt{d_{\text{A}}}\, (\mathbb{1}^{\text{A}}\ot\sqrt{\tau}^{\text{\,B}}) \ket{\phi_+}_{\text{AB}}.
\end{align}
and consider the following choice of operators in (\ref{eq:Fasep}):
\begin{align}
    \tau^{\text{AB}} = \dyad{\tau}_{\text{AB}}, \qquad  O_a^{\text{V}'\text{A}} = d_{\text{V}'} \, \left[ \mathbb{1}^{\text{V}'} \ot (\tau^{-1/2}_{\text{A}})^T  \right] \left(Y_a^{\text{VA}}\right)^T \left[ \mathbb{1}^{\text{V}} \ot (\tau^{-1/2}_{\text{A}})^T  \right].
\end{align}
By plugging these into (\ref{eq:fa}) and performing analogous steps as in (\ref{eq:Ra1}-\ref{eq:Ra2}) we obtain:
\begin{align}
    F_a^{\text{V}\text{B}} &= d_{\text{A}}\, d_{\text{V}'}  \cdot  \tr_{\text{V}'\text{A}}\left[(\mathbb{1}^{\text{V}} \ot Y_a^{\text{V}'\text{A}} \ot \mathbb{1}^{\text{B}}) (\phi_{+}^{\text{V}\text{V}'} \ot \phi_+^{\text{AB}})\right] \\
    &= Y_a^{\text{V}\text{B}}.
\end{align}
Since $Y_a^{\text{V}\text{B}}$ up to this point were arbitrary separable operators satisfying the no-signalling condition, we can infer that any such family of operators can be written as in (\ref{eq:Fasep}). Hence we obtain another equivalence: 
\begin{align}
    \label{eq:equivF}
    \{\Lambda_a^{c}\} \in \mathcal{F} \iff  F_a^{\text{V}\text{B}} \in \sep \quad \text{and} \quad \sum_a  F_a^{\text{V}\text{B}} = \frac{1}{d_{\text{V}}} \, \mathbb{1}^{\text{V}} \ot \tau^{\text{B}}.
\end{align}

Let us now return to the optimization problem (\ref{eq:99}). We multiply both sides of the first line of constraints by $1+r$, label $\widetilde{R}_{a} = r \, R_a$, $\widetilde{F}_a= (1+r) F_a$, $\widetilde{\eta} = r \,\eta$, $\widetilde{\tau} = (1+r) \, \tau$,  and write the second line of constraints using (\ref{eq:equivR}) and (\ref{eq:equivF}).  This allows (\ref{eq:99}) to be written in the equivalent form, which is now manifestly a semi-definite program:
\begin{align}
    \label{eq:211}
    \mathcal{T}(\mathbb{\Lambda}) = \qquad \min_{\{\widetilde{R}_a\},  \{\widetilde{F}_a\}, \widetilde{\tau}, \widetilde{\eta} } \quad & \tr \widetilde{\tau}^{\,\text{B}}-1 \\ \nonumber
    & J_a^{\text{VB}} + \widetilde{R}_a^{\text{VB}} = \widetilde{F}_a^{\text{VB}} \qquad  \qquad  & \forall \, a \\[.3em] \nonumber 
    & \sum_a \widetilde{F}_a^{\text{VB}} = \frac{1}{d_{\text{V}}}\, \mathbb{1}^{\text{V}} \ot \widetilde{\tau}^{\,\text{B}} & \\ \nonumber
    & \sum_a \widetilde{R}_a^{\text{VB}} = \frac{1}{d_{\text{V}}}\, \mathbb{1}^{\text{V}} \ot \widetilde{\eta}^{\,\text{B}} & \\ \nonumber
    & \forall\, a \quad \widetilde{F}_a^{\text{VB}}  \in \sep, \quad \forall\, a \quad \widetilde{R}_a^{\text{VB}}  \geq 0.
\end{align}  

Notice that the first three lines of constraints in (\ref{eq:211}) are linearly dependent, hence w.l.o.g. we skip the third line of constraints. Furthermore,  we can replace the equality in the condition $\sum_a \widetilde{F}_a^{\text{VB}} = d_{\text{V}}^{-1} \cdot \mathbb{1}^{\text{V}} \ot \widetilde{\tau}^{\,\text{B}}$ with an inequality since adding a positive part to $d_{\text{V}}^{-1} \cdot \mathbb{1}^{\text{V}} \ot \widetilde{\tau}^{\,\text{B}}$ can only increase $\tr \widetilde{\tau}^{\,\text{B}}$. Finally, we multiply the second and third line of constraints by $d_{\text{V}}$ (this will lead to a simpler form of the dual problem later on). This allows us to reach the following form of the primal problem:
\begin{align}
    \label{eq:28}
    \mathcal{T}(\Lam{}) = \qquad \min_{\{\widetilde{F}_a^{\text{VB}}\},  \widetilde{\tau}^{\,\text{B}}} \quad & \tr \widetilde{\tau}^{\,\text{B}} - 1  & \qquad\\ \nonumber
    & d_{\text{V}}\cdot \widetilde{F}_a^{\text{VB}} \geq  d_{\text{V}} \cdot J_a^{\,\text{VB}} \qquad \qquad \qquad \forall \, a & \color{gray} A_a^{\text{VB}} \color{black} \\[0.8em] \nonumber
    &d_{\text{V}} \cdot \sum_a \widetilde{F}_a^{\text{VB}} \leq  \mathbb{1}^{\text{V}} \ot  \widetilde{\tau}^{\,\text{B}}, \qquad \forall \, a \quad \widetilde{F}_a^{\text{VB}} \in \sep, & \color{gray} B^{\text{VB}}, \, W_a^{\text{VB}}. \nonumber 
\end{align}
We now look at the dual formulation of the above problem. To do so we first write the associated Lagrangian using the dual variables associated with each set of constraints  (displayed above on the right-hand side in grey): 
\begin{align}
    \mathcal{L} &= \tr \widetilde{\tau}^{\,\text{B}} - 1 - \sum_a \tr A_a^{\text{VB}}\left[d_{\text{V}} \cdot \widetilde{F}_a^{\text{VB}} - d_{\text{V}}\cdot J_a^{\text{VB}}  \right] - \tr B^{\text{VB}}\left[\mathbb{1}^{\text{V}} \ot \widetilde{\tau}^{\,\text{B}} - d_{\text{V}}\cdot \sum_a \widetilde{F}_a^{\text{VB}}\right] - \sum_a \tr\left[W_a^{\text{VB}} \widetilde{F}_a^{\text{VB}}\right] \\
    &= \sum_a \tr \widetilde{F}_a^{\text{VB}}\left[-d_{\text{V}}\cdot A_a^{\text{VB}} + d_{\text{V}} \cdot B^{\text{VB}} - W_a^{\text{VB}} \right] + \tr \widetilde{\tau}^{\,\text{B}}  \left[ \mathbb{1}^{\text{B}} - B^{\text{B}} \right] + d_{\text{V}}\cdot \sum_a \tr \left[A_a^{\text{VB}} J_a^{\text{VB}}\right] - 1.
\end{align}
where $\{A_a^{\text{VB}}\}_a$, $B^{\text{VB}}$ and $\{W_a^{\text{VB}}\}_a$ are the dual variables corresponding to each set of constraints. We can ensure that $\mathcal{L} \leq r$ by ($i$) demanding $A_a^{\text{VB}} \geq 0$, $B^{\text{VB}} \geq 0$ and $W_a^{\text{VB}} \in \mathcal{W}$, where $\mathcal{W} = \{W | \tr[\rho W] \geq 0 \quad \forall \rho \in \mathcal{S} \}$ is, by definition, the set of all entanglement witnesses \cite{Horodecki2009} and ($ii$) demanding that the terms in the square brackets which appear along with the primal variables in the last line vanish. This leads to the following (dual) semi-definite program:
\begin{align}
    \label{eq:27}
    \mathcal{T}(\mathbb{\Lambda}) = \qquad \max_{\{A_a^{\text{VB}}\}_a, \, B^{\text{VB}}} \quad & d_{\text{V}}\cdot \sum_{a} \tr \left[ A_{a}^{\text{VB}} J_{a}^{\text{VB}}\right] - 1  & \qquad \\ \nonumber
    & B^{\text{VB}} - A_a^{\text{VB}} \in \mathcal{W} \qquad\qquad\,\, \forall\, a \\ \nonumber
    & B^{\text{VB}} \geq 0, \quad B^{B} =  \mathbb{1}^{\text{B}}, \qquad \forall \,a\quad A_a^{\text{VB}} \geq 0.
\end{align}
Finally, notice that in our case strong duality holds since we can always find a feasible $\widetilde{F}_a^{\text{VB}} = \alpha \cdot \mathbb{1}^{\text{VB}}$ and $\widetilde{\tau}^{\,\text{B}} = \alpha \cdot \sum_a \mathbb{1}^{\text{B}}$ for some $\alpha \geq 0$ in the primal formulation of the problem. Thus, via the Slater's condition \cite{Boyd2004} we can infer that there is no gap between the solutions of (\ref{eq:28}) and (\ref{eq:27}). 

Let us now return to the primal formulation of the problem (\ref{eq:28}) and let $\widetilde{F}_a^{\text{VB}} = \widetilde{F}_a^*$ and $\widetilde{\tau}^{\,\text{B}} = \widetilde{\tau}^*$ be the optimal choice of primal variables. Notice that $1+\mathcal{T}(\Lam{})  \geq \sum_a  \tr \widetilde{F}_a^*$, where $\widetilde{F}_a^*$ is a separable operator. Denoting $\tau_a^* := \widetilde{F}_a^{*} / \tr \widetilde{F}_a^*$ and $p_S^*(a) := \tr \widetilde{F}_a^* / \left[\sum_{a'}\tr \widetilde{F}_{a'}^*\right]$ we can write:
\begin{align}
    \label{eq:29}
    J_a \leq \widetilde{F}_a^* \leq [1+\mathcal{T}(\Lam{})]\, p_S(a) \cdot \tau_a^*,
\end{align}
where $\tau_a^*$ is a separable state and $p_S(a)$ forms a probability distribution. Equivalently, we can write this inequality in terms of Choi-Jamiołkowski operators as: $J_a \leq [1+\mathcal{T}(\Lam{})] F_a$, where $\{F_a\}$ is a set of Choi-Jamiołkowski operators corresponding to a classical teleportation instrument $\Lam{}^c = \{\Lambda_a^c\}$. 

\section{Properties of RoT}
In this Appendix we prove the three properties of robustness of teleportation highlighted in the main text. 
\paragraph{Faithfulness} 
If a teleportation instrument is classical, that is $\Lam{} \in \mathcal{F}$, then we can always choose a feasible $r = 0$ in the defining optimization problem (\ref{eq:5}). Since $\mathcal{T}(\Lam{})$ is non-negative, then $r = 0$ is also optimal.
\paragraph{Convexity} Let $\{ \Lambda_a'^{(1)}[\omega_x],\, \Lambda_a^{c \, (1)}[\omega_x]\}$ be the optimal primal variables in the defining problem  (\ref{eq:5}) for $\mathcal{T}(\Lam_1)$ with $\Lam{}_1 = \{\Lambda_a^{(1)}\}$ and similarly for $\{ \Lambda_a'^{(2)}[\omega_x],\, \Lambda_a^{c \, (2)}[\omega_x]\}$ and $\mathcal{T}(\Lam_2)$ with $\Lam{}_2 = \{\Lambda_a^{(2)}\}$. Let $\Lam{}' = \{\Lambda'_a \}_a$ be a convex mixture of the two teleportation instruments, that is $\Lambda_a'[\cdot] = p \, \Lambda_a^{(1)}[\cdot] + (1-p) \, \Lambda_a^{(2)}[\cdot]$ for each $a$. We can construct (potentially sub-optimal) solutions for $\mathcal{T}(\Lam{}')$ using: $\Lambda_a'[\omega_x] = p \, \Lambda_a'^{(1)}[\omega_x] + (1-p)\, \Lambda_a'^{(2)}[\omega_x]$ and $\Lambda_a^c[\omega_x]' = p \, \Lambda_a^{c \, (1)}[\omega_x] + (1-p)\, \Lambda_a^{c \, (2)}[\omega_x]$. Substituting $\Lambda_a'[\omega_x]'$ and $\Lambda_a^c[\omega_x]'$ into the constraints of problem (\ref{eq:5}) for $\Lam{}'$ shows that this choice is feasible. This leads to the upper bound on $\mathcal{T}(\Lam{}')$: 
\begin{align}
    \mathcal{T}(\Lam{}') \leq \tr \sum_a \Lambda_a'[\omega_x]' = p\, \cdot \tr \sum_{a} \Lambda_a'^{(1)}[\omega_x] + (1-p)\, \cdot \tr \sum_{a} \Lambda_a'^{(2)}[\omega_x] = p \cdot \mathcal{T}(\Lam{}_1) + (1-p)\cdot \mathcal{T}(\Lam{}_2).
\end{align}
\paragraph{Monotonicity} Let us start with quantum simulation. Assume that $\Lam{}$ can simulate $\Lam{}'$, i.e. $\Lam{} \succ_q \Lam{}'$. This means that there exists a collection of channels $\Theta_{\lambda}$, $\Omega_{\lambda}$ and probability distributions $p_{\lambda}$ and $p(b|a, \lambda)$ such that for all $b$:
\begin{align}
    \Lambda_b'(\cdot) = \sum_{a, \lambda} p_{\lambda} p(b|a, \lambda) \Theta_{\lambda} \circ \Lambda_a' \circ \Omega_{\lambda} (\cdot)
\end{align}
Suppose now that we solved the dual problem (\ref{eq:27}) for $\mathcal{T}(\Lam{}')$ using optimal dual variables $B'$ and $A_b'$. Using these we can construct an educated guess for  $\mathcal{T}(\Lam{}')$ in the following way:
\begin{align}
    \label{eq:22}
    B^* = \sum_{b, \lambda} p_{\lambda} \, p(b|a, \lambda)\, \left((\Omega^{T}_{\lambda})^{\dagger} \ot  \Theta_{\lambda}^{\dagger}\right) [B], \qquad A_a^* = \sum_{b, \lambda} p_{\lambda} \, p(b|a, \lambda)\, \left((\Omega^{T}_{\lambda})^{\dagger} \ot \Theta_{\lambda}^{\dagger}\right) [A_b'].
\end{align}
Using these we can find the following lower bound:
\begin{align}
    1 + \mathcal{T}(\Lam{}) &\geq d_{\text{V}} \cdot \sum_{a} \tr[J_a \cdot A_a^*] \\
    &= d_{\text{V}} \cdot \sum_{a, b, \lambda} p_{\lambda} \, p(b|a,\lambda)\, \tr[(\mathcal{I} \ot \Lambda_a) \, [\phi_+] \cdot \left((\Omega^{T}_{\lambda})^{\dagger} \ot \Theta_{\lambda}^{\dagger}\right) [A_b']] \\
    &= d_{\text{V}} \cdot \sum_{a, b, \lambda} p_{\lambda} \, p(b|a,\lambda) \, \tr[(\Omega^{T}_{\lambda} \ot \Lambda_a)\, [\phi_+] \cdot (\mathcal{I} \ot \Theta_{\lambda}^{\dagger})\,  [A_b']] \\
    &= d_{\text{V}} \cdot \sum_{a, b, \lambda} p_{\lambda} \, p(b|a,\lambda) \, \tr[(\mathcal{I} \ot \Lambda_a \circ \Omega_{\lambda}) [\phi_+] \cdot (\mathcal{I} \ot \Theta_{\lambda}^{\dagger}) \, [A_b']]\\
    &= d_{\text{V}} \cdot \sum_{b} \tr[(\mathcal{I} \ot \Lambda_b')\, [\phi_+] \cdot A_b'] \\
    &= 1 + \mathcal{T}(\Lam{}').
\end{align}
Let us now show that the choice (\ref{eq:22}) is feasible. By construction we have $B^* \geq 0$, $A_a^* \geq 0$ and $\tr_{\text{V}} B^* = \mathbb{1}$, since:
\begin{align}
    \tr_{\text{V}} \left[ \left( (\Omega_{\lambda}^{T})^{\dagger} \ot \Theta_{\lambda}^{\dagger}\right) B\right] &=  \tr_{\text{V}} \left[ \left( (\Omega_{\lambda}^{T})^{\dagger} \ot \Theta_{\lambda}^{\dagger} \circ \mathcal{B}^{\dagger} \right) [\phi_+] \right] \\
    &= \tr_{\text{V}} \left[ \left( \mathcal{I} \ot \Theta_{\lambda}^{\dagger} \circ \mathcal{B}^{\dagger} \circ \Omega_{\lambda}^{\dagger}\right) [\phi_+]\right] \\
    &= \Theta_{\lambda}^{\dagger} \circ \mathcal{B}^{\dagger} \circ \Omega_{\lambda}^{\dagger}(\mathbb{1}) \\
    &= \mathbb{1},
\end{align}
where in the first line we used the Choi-Jamiołkowski isomorphism ${B} = (\mathcal{I}\ot \mathcal{B}^{\dagger}) [\phi_+]$ for a map $\mathcal{B} \in \cptp$ and in the third line we used the fact that the adjoint of a CPTP map is unital. It remains to show that $B^* - A_a^*$ is an entanglement witness. Let $\rho_S$ be an arbitrary separable state. We have:
\begin{align}
    \tr[(B^* - A_a^*) \rho_S] &= \sum_{b, \lambda} p_{\lambda}\, p(b|a, \lambda) \, \tr\left[\left((\Omega^{T}_{\lambda})^{\dagger} \ot  \Theta_{\lambda}^{\dagger}\right) [B - A_b'] \cdot \rho_S\right] \\
    &= \sum_{b, \lambda} p_{\lambda}\, p(b|a, \lambda) \, \tr[(B - A_b') \cdot \left(\Omega^{T}_{\lambda} \ot  \Theta_{\lambda}\right) [\rho_S]]  \\
    &= \sum_{b, \lambda} p_{\lambda}\, p(b|a, \lambda) \, \tr[(B - A_b') \cdot \rho_{\lambda}']  \\
    &\geq 0,
\end{align}
where we used the fact that $W_b = B - A_b'$ is by assumption an entanglement witness and $\rho_{\lambda}' = \left(\Omega^{T}_{\lambda} \ot  \Theta_{\lambda}\right) [\rho_S]$ is a separable operator. To show analogous statement about classical simulation is simple, as this is just a special case of quantum simulation resulting from choosing $p_{\lambda} = \frac{1}{o_{\lambda}}$, where $o_{\lambda}$ size of the alphabet associated with $\lambda$, $\Theta_{\lambda} = \Omega_{\lambda} = \mathcal{I}$.

\section{RoT as an advantage in the teleportation of quantum correlations}
Here we prove that the robustness of teleportation $\mathcal{T}(\mathbb{\Lambda})$ can be viewed as the best advantage in the task of teleporting quantum correlations using a fixed quantum teleportation instrument $\mathbb{\Lambda}$ over any classical teleportation instrument. We start by constructing a particular game $\mathcal{G}^*$ using the dual formulation of the RoT and then show that $1 + \mathcal{T}(\Lam{})$ gives a meaningful lower bound on the advantage. We then use primal formulation (\ref{eq:28}) and show that $1 + \mathcal{T}(\Lam{})$ also bounds the advantage from above. 

Let us begin by noting that the classical average score $q^c(\mathcal{G}) $ for game $\mathcal{G}$ is given by:
\begin{align}
    q^c(\mathcal{G}) = \max_{\Lam{}^c \in \mathcal{F}}\, \max_{\{\mathcal{U}_a\} \in \unitary} \, \sum_a  f(a) \tr \left[ (\mathcal{I}\ot \mathcal{U}_a \circ \Lambda_a^c) \, [\sigma] \cdot \xi_a \right].
\end{align}
Suppose we have solved the dual problem for the RoT as given by (\ref{eq:27}) using dual variables $B$ and $A_a$. We can construct a (potentially sub-optimal) task $\mathcal{G}^* = \{\sigma^*, \xi^*_a, f^*(a)\}$ using these optimal variables in the following way:
\begin{align}
    & \sigma^* = \phi_+, \quad \xi_a^* = \frac{A_a}{\tr A_a}, \quad f^*(a) = {\tr A_a}. 
\end{align}
The maximal average score which can be achieved using classical teleportation instruments $\mathbb{\Lambda}^c$ in game $\mathcal{G}^*$ can be bounded by:
\begin{align}
    q^c(\mathcal{G}^*) = \max_{\mathbb{\Lambda}^c \in \mathcal{F}} \, q(\mathcal{G}^*, \mathbb{\Lambda}^c) &=  \max_{\mathbb{\Lambda}^c \in \mathcal{F}} \max_{\{\mathcal{U}_a\} \in \unitary } \, \sum_{a} f^*(a) \tr \left[ (\mathcal{I}\ot \mathcal{U}_a \circ \Lambda_a^c) \, [\sigma^*] \cdot \xi_a^* \right]
    \\ &=  \max_{\mathbb{\Lambda}^c \in \mathcal{F}} \max_{\{\mathcal{U}_a\} \in \unitary } \, \sum_{a} \tr \left[ (\mathcal{I}\ot  \mathcal{U}_a \circ \Lambda_a^c) \, [\phi_+] \cdot A_a \right]
    \\  &=  \max_{\{F_a\} \in \mathcal{F}} \max_{\{\mathcal{U}_a\} \in \unitary }\, \sum_{a}  \tr \left[(\mathcal{I}\ot \mathcal{U}_a)[F_a] \cdot (B - W_a) \right] \\
    &\leq \max_{\{F_a\} \in \mathcal{F}} \, \sum_{a} \tr \left[F_a \cdot B \right]  \\ 
    & = \frac{1}{d_{\text{V}}} \max_{\tau^{\text{B}}} \,  \tr \left[\tau^{\text{B}} \cdot B^{\text{B}} \right]  \\
    &\leq \frac{1}{d_{\text{V}}},
\end{align}
where in the second line we used $F_a = (\mathcal{I}\ot \Lambda_a^c) \,\phi_+ \in \sep$ and with a slight abuse of notation we denoted optimization over Choi-Jamiołkowski operators $F_a$ corresponding to classical teleportation instruments as $\{F_a\} \in \mathcal{F}$. In the third line we used the constraint from the dual: $B - A_a = W_a \in \mathcal{W}$ and in the fourth line we employed the fact that $W_a$ is an entanglement witness. Finally, in fifth line we used the no-signalling condition $\sum_a F_a = d_{\text{V}}^{-1} \cdot \mathbb{1}^{\text{V}} \ot \tau^{\text{B}}$.

Notice now that for an arbitrary teleportation instrument $\Lam{} = \{\Lambda_a\}$ with Choi-Jamiołkowski operators $J_a = (\mathcal{I}\ot\Lambda_a)[\phi_+]$ we can write:
\begin{align}
    \label{eq:12}
    \max_{\mathcal{G}} \frac{q(\mathcal{G}, \mathbb{\Lambda})}{ q^c(\mathcal{G})} \geq \frac{q(\mathcal{G}^*, \mathbb{\Lambda})}{q^c(\mathcal{G}^*)} \geq  d_{\text{V}} \sum_a \tr \left[J_a \cdot A_a \right] = 1 + \mathcal{T}(\Lam{}).
\end{align}
To prove the reverse direction let $d = d_{\text{A}'} = d_{\text{B}'}$ and notice that using (\ref{snake_id}) we can rewrite any bipartite state $\sigma^{\text{AB}}$ as:
\begin{align}
    \label{eq:sigma_rew}
    \sigma^{\text{AB}} = d^2 \, \tr_{\text{A}'\text{B}'}\left[(\mathbb{1}^{\text{A}} \ot \phi_+^{\text{A}'\text{B}'} \ot \mathbb{1}^{\text{B}}) (\sigma^{\text{AA}'} \ot \phi_+^{\text{B}'\text{B}}) \right].
\end{align}
Let $\Lam{} = \{\Lambda_a^{\text{B}\rightarrow\text{C}}\}$ be an arbitrary teleportation instrument with Choi-Jamiołkowski operators  $\{J_a\}$. Then (\ref{eq:sigma_rew}) leads to:
\begin{align}
    (\mathcal{I}^{\text{A}} \ot \Lambda_a^{\text{B} \rightarrow \text{C}}) [\sigma^{\text{AB}}] &= d^2 \, \tr_{\text{A}'\text{B}'}\left[(\mathbb{1}^{\text{A}} \ot \phi_+^{\text{A}'\text{B}'} \ot \mathbb{1}^{\text{B}}) (\sigma^{\text{AA}'} \ot  (\mathcal{I}^{\text{B}'} \ot \Lambda_a^{\text{B} \rightarrow \text{C}})[\phi_+^{\text{B}'\text{B}}]) \right] \\
    &= d^2 \, \tr_{\text{A}'\text{B}'}\left[(\mathbb{1}^{\text{A}} \ot \phi_+^{\text{A}'\text{B}'} \ot \mathbb{1}^{\text{B}}) (\sigma^{\text{AA}'} \ot J_a^{\text{B}'\text{C}}) \right] \\
    &\leq  d^2 \, (1+\mathcal{T}(\Lam{})) \tr_{\text{A}'\text{B}'}\left[(\mathbb{1}^{\text{A}} \ot \phi_+^{\text{A}'\text{B}'} \ot \mathbb{1}^{\text{B}}) (\sigma^{\text{AA}'} \ot {F}_a^{\text{B}'\text{C}}) \right] \\
    &= (1+\mathcal{T}(\Lam{})) \cdot (\mathcal{I}^{\text{A}} \ot (\Lambda_a^c)^{\text{B} \rightarrow \text{C}}) [\sigma^{\text{AB}}] \label{eq:choi_upperbnd}
\end{align}
where in the second line we used (\ref{eq:29}) to upper-bound $J_a$ with the Choi-Jamiołkowski operator $F_a$ of a classical teleportation instrument $\{\Lambda_a^c\}$. We can now calculate the average score for an arbitrary game $\mathcal{G} = \{\sigma, \xi_a, f(a)\}$:
\begin{align}
    q(\mathcal{G}, \mathbb{\Lambda}) &=  \, \max_{{\Lam{}' \leq \Lam{}}}\, \max_{\{\mathcal{U}_a\} \in \unitary} \, \sum_{a} f(a) \tr \left[ (\mathcal{I}\ot \mathcal{U}_a \circ \Lambda_a') \, [\sigma] \cdot \xi_a \right] \\
    & \leq \max_{\mathbb{\Lambda}' \leq \mathbb{\Lambda}} \, \max_{\{\mathcal{U}_a\} \in \unitary} \, [1 + \mathcal{T}(\mathbb{\Lambda}')] \, \sum_a  f(a) \tr \left[ (\mathcal{I}\ot \mathcal{U}_a \circ \Lambda_a^c) \, [\sigma] \cdot \xi_a \right] \\
    &\leq \max_{\mathbb{\Lambda}' \leq \mathbb{\Lambda}} \,  [1 + \mathcal{T}(\mathbb{\Lambda}')] \,
    \max_{\Lam{}^c \in \mathcal{F}}\, \max_{\{\mathcal{U}_a\} \in \unitary} \, \sum_a  f(a) \tr \left[ (\mathcal{I}\ot \mathcal{U}_a \circ \Lambda_a^c) \, [\sigma] \cdot \xi_a \right] \\
    & = \max_{\mathbb{\Lambda}' \leq \mathbb{\Lambda}}\, [1 + \mathcal{T}(\mathbb{\Lambda}')] \,\,  q^c (\mathcal{G}) \\
    & \leq [1 + \mathcal{T}(\mathbb{\Lambda})] \, q^c (\mathcal{G}),
\end{align}
where in the first line we used (\ref{eq:choi_upperbnd}) and in the fourth line we used the monotonicity property of the RoT (\ref{eq:18}). Note that the above reasoning is valid for any game $\mathcal{G}$ and thus by taking the maximum over all $\mathcal{G}$ we obtain:
\begin{align}
    \max_{\mathcal{G}} \frac{q(\mathcal{G}, \mathbb{\Lambda})}{ q^c (\mathcal{G})} \leq 1 + \mathcal{T}(\mathbb{\Lambda}).
\end{align}
Combined with the lower bound, this proves the equality.

\section{RoT as an advantage in subchannel discrimination with quantum side information}
Let $\mathbb{E} = \{\mathcal{E}_x\}$ be an instrument, such that $\sum_x \mathcal{E}_x[\cdot] = \mathcal{E}[\cdot]$ forms a valid quantum channel, and let $\mathcal{A} = \{\{M_a\}, \, \rho \}$ be a resource used in the game; it consists of a bipartite measurement $\{M_a\} \in \povm$ and a bipartite state $\rho$ acting as a quantum memory in the game. The average probability of guessing which subchannel from $\mathbb{E}$ was applied locally to $\rho$ is given by:
\begin{align}
    \label{eq:24}
    p_{\text{succ}}(\mathbb{E}, \mathcal{A}) = \max_{p(g|a)}\, \sum_{x,a,g} p(g|a)  \, \tr[(\mathcal{E}_x \ot \mathcal{I}) [\rho] \cdot M_a ] \, \delta_{g, x}.
\end{align}
In what follows we will use the following operator identity:
\begin{align}
    \label{eq:25}
    d_{\text{V}}^2 \, \tr[X^{\text{VB}}\, \phi_+^{\text{VB}}]  = \tr \left[ (\mathcal{E}^{\text{A}} \ot \mathcal{I}^{\text{B}})  [\rho^{\text{AB}}] \cdot M^{\text{AB}} \right],
\end{align}
where:
\begin{align}
X^{\text{VB}} = \tr_{\text{V}'\text{A}} \left[ \left(\mathbb{1}^{\text{V}} \ot M^{\text{V}'\text{A}} \ot \mathbb{1}^{\text{B}}\right)\cdot \left( (\mathcal{I}^{\text{V}} \ot \mathcal{E}^{\text{V}'}) \, [\phi_+^{\text{V}\text{V}'}] \ot \rho^{\text{AB}} \right) \right]    
\end{align}
and $\mathcal{E}$ is an arbitrary channel. The above identity can be proven by direct substitution and becomes almost natural when expressed in a diagrammatic form \cite{Coecke2017,Wood2011,Biamonte2017}. 

Let us now use the identity (\ref{eq:25}) and recall  that $\Lambda_a^{\text{V}' \rightarrow \text{B}}$ is a subchannel from the teleportation instrument $\Lam{} = \{\Lambda_a\}$ which was defined in (\ref{eq:telep_instr}) and acts in the following way:
\begin{align}
    \Lambda_a^{\text{V}'\rightarrow \text{B}}[\omega] = \tr_{\text{V}'\text{A}} [(M_a^{\text{V}'\text{A}} \ot \mathbb{1}^{\text{B}})(\omega^{\text{V}'} \ot \rho^{\text{AB}})]
\end{align}
The associated Choi–Jamiołkowski operators $J_a^{\,\text{VB}} = (\mathcal{I}^{\text{V}} \ot \Lambda_a^{\text{V}' \rightarrow \text{B}}) \, [\phi_+^{\text{VV}'}]$ are given by:
\begin{align}
    J_a^{\text{VB}} = \tr_{\text{V}'\text{A}} [(\mathbb{1}^{\text{V}} \ot M_a^{\text{V}'\text{A}} \ot \mathbb{1}^{\text{B}})(\phi_+^{\text{V}\text{V}'} \ot \rho^{\text{AB}})]
\end{align}
This leads to the following realization:
\begin{align}
    d_{\text{V}}^2\cdot \tr[(\mathcal{I}^{\text{V}} \ot \mathcal{E}_x^{\text{B}})  \left[J_a^{\text{VB}}\right] \cdot \phi_+^{\text{VB}}] &= d_{\text{V}}^2\cdot \tr[((\mathcal{E}_x^{\text{V}})^{T}\ot\mathcal{I}^{\text{B}})  \left[J_a^{\text{VB}}\right] \cdot \phi_+^{\text{VB}}]
    \\ &= d_{\text{V}}^2\cdot \tr[((\mathcal{E}_x^{\text{V}})^{T}\ot\mathcal{I}^{\text{B}})  \left[ \tr_{\text{V}'\text{A}} [(\mathbb{1}^{\text{V}} \ot M_a^{\text{V}'\text{A}} \ot \mathbb{1}^{\text{B}})(\phi_+^{\text{V}\text{V}'} \ot \rho^{\text{AB}})] \right] \cdot \phi_+^{\text{VB}}] 
    \\ &= d_{\text{V}}^2 \cdot \tr\left[\tr_{\text{V}'\text{A}} \left[ \left(\mathbb{1}^{\text{V}} \ot M^{\text{V}'\text{A}} \ot \mathbb{1}^{\text{B}}\right) \left( (\mathcal{I}^{\text{V}} \ot \mathcal{E}^{\text{V}'}) \, [\phi_+^{\text{V}\text{V}'}] \ot \rho^{\text{AB}} \right) \right] \cdot \phi_+^{\text{VB}}\right]  \\
    &=\tr \left[ (\mathcal{E}_x^{\text{A}} \ot \mathcal{I}^{\text{B}} )  [\rho^{\text{AB}}] \cdot M_a^{\text{AB}}  \right],
\end{align}
where in the first and the third line we made use of the special property of the maximally-entangled state (\ref{eq_prop_ment}). In this way we can rewrite (\ref{eq:24}) as:
\begin{align}
    p_{\text{succ}}(\mathbb{E}, \mathcal{A}) &= d_{\text{V}}^2 \cdot \max_{p(g|a)}\, \sum_{x,a,g}  p(g|a)  \, \tr[(\mathcal{I} \ot \mathcal{E}_x)  [J_a] \cdot \phi_+] \, \delta_{g, x} \\
    &= d_{\text{V}}^2 \cdot \max_{p(x|a)}\, \sum_{x,a}  p(x|a)  \, \tr[J_a \cdot (\mathcal{I} \ot \mathcal{E}_x^{\dagger})[\phi_+]] \label{eq:34}
\end{align}

Suppose now that we have solved the dual problem for the RoT as given by (\ref{eq:27}) using dual variables $B^*$ and $A_x^*$. Using  these optimal variables we will now construct a sequence of games $\mathbb{E}^* = \{\mathcal{E}_x^* \}$, parametrized with $N$, that is the number of subchannels forming the instrument. This proof technique is inspired by the methods used in \cite{Piani2015}.  Let us define a set of subchannels via their duals, i.e:
\begin{align}
    (\mathcal{E}_x^{*})^{\dagger}[\rho] =
    \begin{dcases}
    \alpha \, \tr_{\text{V}} \left[(\rho^T \ot \mathbb{1}) \, A_x^*\right] \qquad &\text{for} \qquad 1 \leq x \leq o_a, \\
    \frac{1}{N \cdot d_{\text{V}}} \left[\mathbb{1} - \alpha \sum_{x' = 1}^{o_a} (A_{x'}^*)^\text{B}  \right] \tr[\rho] \qquad &  \text{for} \qquad o_a + 1 \leq x \leq o_a + N.
    \end{dcases}
\end{align}
In the above $\alpha = \norm{\sum_{x' = 1}^{o_a} A_{x'}^B }_{\infty}^{-1}$ is a real parameter chosen such that the map defined above is completely positive.  Notice that the constraints of the dual problem (\ref{eq:27}) imply that $0 \leq A_{x'} \leq \mathbb{1}$. To verify that $\mathcal{E}^{*} = \sum_{x} \mathcal{E}_x^*$ defines a channel recall that $\mathcal{E} \in \cptp$ if and only if its dual map $\mathcal{E}^{\dagger}$ is unital. By construction we have:
\begin{align}
    \sum_{x=1}^{o_a + N} (\mathcal{E}_x^{*})^{\dagger}[\mathbb{1}] = \alpha \sum_{x=1}^{o_a} \tr_{\text{V}}\left[ A_x^*\right] + \mathbb{1} - \alpha \sum_{x=1}^{o_a} \tr_{\text{V}}\left[ A_x^*\right]  = \mathbb{1}.
\end{align}
Notice that by our particular definition of the instrument $\mathcal{G}^*$ we also have the following relation:
\begin{align}
        \left[\mathcal{I} \ot (\mathcal{E}_x^{*})^{\dagger}\right] [\phi_+] =
        \begin{dcases}
        \frac{\alpha}{d_{\text{V}}} A_x^* \qquad &\text{for} \qquad 1 \leq x \leq o_a, \\
        \frac{1}{N\cdot d_{\text{V}}^2}\, \mathbb{1} \ot \left( \mathbb{1} - \alpha \sum_{x'=1}^{o_a} \tr_{\text{V}} \left[A_{x}^*\right] \right) \qquad &  \text{for} \qquad o_a + 1 \leq x \leq o_a + N.
        \end{dcases}
\end{align}
Let us now upper bound the maximal probability of guessing in a game specified by $\mathcal{G}^*$ and when having access only to classical resources. This is specified by $p_{\text{succ}}^c(\mathbb{E}^*) = \max_{\mathcal{A}^c \in \mathcal{F}} \, p_{\text{succ}}(\mathbb{E}^*, \mathcal{A}^c)$, where the optimization is performed over all $\mathcal{A}^c = \{\{M_a\}, \sigma \}$ with $\sigma \in \sep$ and arbitrary measurements $\{M_a\}$. Using (\ref{eq:34}) and the fact that this optimization is equivalent to an optimization over a classical teleportation instrument with Choi-Jamiołkowski operators $F_a = (\mathcal{I} \ot \Lambda_a^c) \phi_+ = p_{\text{T}}(a) \, \sigma_a$ for some $\sigma_a \in \sep$ and probability distribution $p_{\text{T}}(a)$, this becomes:
\begin{align}
    \label{eq:26}
    p_{\text{succ}}^c(\mathbb{E}^*) &=d_{\text{V}}^2  \max_{\substack{\sigma_a \in \sep, \, p_{\text{T}}(a)}} \, \max_{p(x|a)} \, \sum_{x,a}  p(x|a)\, p_{\text{T}}(a) \tr[(\mathcal{I} \ot \mathcal{E}_x^*) [\sigma_a] \cdot \phi_+]  \\ \nonumber
    &= d_{\text{V}}^2  \max_{\substack{\sigma_a \in \sep, \, p_{\text{T}}(a)}}  \max_{p(x|a)}  \sum_a p_{\text{T}}(a)\! \left[ \frac{\alpha}{d_{\text{V}}} \sum_{x=1}^{o_a} p(x|a)    \tr[\sigma_a A_x^*] + \frac{1}{N \cdot d_{\text{V}}^2} \sum_{x=o_a+1}^{o_a+N} p(x|a)  \tr[\sigma_a - \alpha \left(\mathbb{1} \ot \sum_{x'=1}^{o_a} (A_{x'}^*)^B \right) \sigma_a]\right] \\ \nonumber
    &  \leq  d_{\text{V}}  \max_{\substack{\sigma_a \in \sep, \, p_{\text{T}}(a)}} \max_{p(x|a)}   \sum_a p_{\text{T}}(a)\! \left[  \alpha \sum_{x=1}^{o_a} p(x|a)  \tr[\sigma_a A_x^*] + \frac{1}{N \cdot d_{\text{V}}} \sum_{x=o_a+1}^{o_a+N} p(x|a)  \tr[\sigma_a]\right] \\ \nonumber
    &\leq d_{\text{V}} \max_{\substack{\sigma_a \in \sep, \, p_{\text{T}}(a)}} \max_{p(x|a)}  \sum_a p_{\text{T}}(a)\! \left[  \alpha \sum_{x=1}^{o_a} p(x|a) \tr[\sigma_a A_x^*]\right] + \frac{1}{N}.
\end{align}
In the third line we used the fact that the subchannels corresponding to fictitious outcomes $o_a+1\leq x\leq o_a+N$ are positive. Recall that the operators $A_x^*$ must satisfy certain constraints in order to be feasible solutions of the dual problem (\ref{eq:27}). In particular, $A_x^* = B^* - W_x^*$, where $B^*$ is a positive matrix with $\tr_{\text{V}} B^* = \mathbb{1}$ and $W_x^* \in \mathcal{W}$ is an entanglement witness. This allows for the following bound to be obtained:
\begin{align}
    \sum_a p_{\text{T}}(a)  \sum_{x=1}^{o_a} p(x|a) \tr[\sigma_a A_x^*] &= \sum_a p_{\text{T}}(a)  \sum_{x=1}^{o_a} p(x|a) \tr[\sigma_a (B^*-W_x^*)] \\
    &\leq \sum_a p_{\text{T}}(a)  \sum_{x=1}^{o_a} p(x|a) \tr[\sigma_a B^*] \\
    &\leq \sum_a p_{\text{T}}(a) \tr[\sigma_a B^*] \\
    &= \frac{1}{d_{\text{V}}} \tr[(\mathbb{1} \ot \sigma^{\text{B}}) B^*] \\
    &= \frac{1}{d_{\text{V}}} \tr[\sigma^{\text{B}}(B^*)^{\text{B}}] \\
    &= \frac{1}{d_{\text{V}}}.
\end{align}
In the first line we used the fact that for a separable $\sigma_a$ and entanglement witness $W_a^*$ the value of $\tr[\sigma_a W_x^*]$ is always positive. In the third line we used the fact that $p_{\text{T}}(a)$ and $\sigma_a$ is an ensemble arising from a (classical) teleportation instrument and thus it satisfies the no-signalling condition, i.e. $\sum_a p_{\text{T}}(a) \, \sigma_a = \frac{1}{d_{\text{V}}} \mathbb{1} \ot \sigma$ for some state $\sigma$. These realizations lead to the following bound on the classical probability of guessing (\ref{eq:26}) in game $\mathbb{E}^*$:
\begin{align}
    \label{eq:35}
    p_{\text{succ}}^c(\mathbb{E}^*) &\leq \alpha + \frac{1}{N}.  
\end{align}
Let us now bound the average probability of guessing in game $\mathbb{E}^*$ when having access to a resource $\mathcal{A}$. We have:
\begin{align}
     p_{\text{succ}}(\mathbb{E}^*, \mathcal{A})  &=
     d_{\text{V}}^2 \cdot \max_{p(x|a)} \, \sum_{x,a} p(x|a)\, p_{\text{T}}(a) \tr[(\mathcal{I} \ot \mathcal{E}_x^*) [\rho_a] \cdot \phi_+] \\
     &\geq \alpha\, d_{\text{V}} \cdot \sum_{a} p_{\text{T}}(a) \tr[\rho_a A_a^* ] \\
     &= \alpha \cdot \left[1 + \mathcal{T}(\Lam{})\right]. \label{eq:30}
\end{align}
In the second line we chose a strategy which does not use the fictitious outcomes, i.e. $p(x|a) = \delta_{x,a}$ and used the identity: $[\mathcal{I} \ot (\mathcal{E}_x^*)^{\dagger}][\phi_+] = \frac{\alpha}{d_{\text{V}}}A_x^*$. Combining bounds (\ref{eq:35}) and (\ref{eq:30}) we find that the maximal advantage optimized over all games is lower bounded by:
\begin{align}
    \label{eq:31}
    \max_{\mathbb{E}^*} \, \frac{p_{\text{succ}}(\mathbb{E}, \mathcal{A})}{p_{\text{succ}}^c(\mathbb{E})} \geq \frac{p_{\text{succ}}(\mathbb{E}^*, \mathcal{A})}{p_{\text{succ}}^c(\mathbb{E}^*)} \geq \left[1 + \mathcal{T}(\Lam{})\right] \cdot \frac{1}{1+\frac{1}{\alpha N}},
\end{align}
where $\Lam{}$ is a teleportation instrument constructed from $\mathcal{A}$. Since we are free to choose $N$ as big as we like, in the limit $N \rightarrow \infty$ the advantage is lower-bounded by $1 + \mathcal{T}(\Lam{})$. To prove the reverse direction we look at the probability of guessing for an arbitrary game $\mathbb{E}$:
\begin{align}
    p_{\text{succ}}(\mathbb{E}, \mathcal{A}) &=    d_{\text{V}}^2 \cdot \max_{p(g|a)} \sum_{x,a,g} p(g|a)  \, \tr[(\mathcal{I} \ot \mathcal{E}_x)[J_a] \cdot \phi_+] \, \delta_{g, x} \\
    &\leq [1 + \mathcal{T}(\Lam{})] \, d_{\text{V}}^2 \cdot \max_{p(g|a)}\, \sum_{x,a,g} p(g|a)\, p_{\text{T}}(a)  \tr[( \mathcal{I} \ot \mathcal{E}_x ) \, [\sigma_a] \cdot \phi_+] \, \delta_{g, x} \\
     &\leq [1 + \mathcal{T}(\Lam{})] \, d_{\text{V}}^2 \cdot  \max_{\sigma_a \in \sep,\, p_{\text{T}}(a)}\, \max_{p(g|a)}\, \sum_{x,a,g} p(g|a) \, p_{\text{T}}(a) \tr[(\mathcal{I} \ot \mathcal{E}_x ) \, [\sigma_a] \cdot \phi_+] \, \delta_{g, x} \\
     &= [1 + \mathcal{T}(\Lam{})] \, p_{\text{succ}}^c(\mathcal{G}),
\end{align}
where the first inequality follows from (\ref{eq:29}), that is $J_a \leq [1+\mathcal{T}(\Lam{})] \, p_{\text{T}}(a)\, \sigma_a$ for a probability distribution $p_{\text{T}}(a)$ and a separable state $\sigma_a$. Since this holds for any game $\mathbb{E}$ we can equivalently write:
\begin{align}
    \label{eq:32}
    \max_{\mathbb{E}} \, \frac{p_{\text{succ}}(\mathbb{E}, \mathcal{A})}{p_{\text{succ}}^c(\mathbb{E})} \leq 1 + \mathcal{T}(\Lam{}).
\end{align}
Combining the bounds (\ref{eq:31}) (in the limit $N\rightarrow \infty$)  and (\ref{eq:32}) we arrive at:
\begin{align}
    \max_{\mathbb{E}} \frac{p_{\text{succ}}(\mathbb{E}, \mathcal{A})}{p_{\text{succ}}^c(\mathbb{E})} = 1 + \mathcal{T}(\Lam{}).
\end{align}
Notice that so far our choice for $\alpha$ was somewhat arbitrary. In order to find its physical interpretation let us consider the maximal probability of guessing using classical resources $\mathcal{A}^c = \{M_a, \sigma\}$:
\begin{align}
    p_{\text{succ}}^c(\mathbb{E}) &= \max_{\mathcal{A}^c \in \mathcal{F}} \max_{p(g|a)}\, \sum_{x,a,g} p(g|a)  \, \tr[(\mathcal{E}_x \ot \mathcal{I}) [\sigma] \cdot M_a] \, \delta_{g, x} \\
    &= \max_{\sigma \in \sep}\, \max_{\{M_a\} \in \povm}\, \max_{p(g|a)}\, \sum_{x,a,g} p(g|a)  \, \tr[(\mathcal{E}_x \ot \mathcal{I}) [\sigma] \cdot M_a] \, \delta_{g, x}
\end{align}
Notice that due to the convex structure of the set of all separable states w.l.o.g we can assume that the optimal separable state $\sigma$ is of the product form $\sigma = \omega \ot \omega'$. This allows to write:
\begin{align}
    p_{\text{succ}}^c(\mathbb{E}) &= \max_{\omega, \, \omega'} \, \max_{\{M_a\} \in \povm} \,\max_{p(x|a)}\, \sum_{x,a} p(x|a)  \, \tr[(\mathcal{E}_x[\omega] \ot \omega') \cdot M_a] \\
    &= \max_{\omega}  \, \max_{\{M_a'\} \in \povm} \, \max_{p(x|a)}\, \sum_{x,a} p(x|a)  \, \tr [\mathcal{E}_x[\omega]\cdot M_a'] \\
    &= \max_{\omega}  \, \max_{\{M_x''\}\in\povm} \, \sum_{x}  \, \tr [\mathcal{E}_x[\omega] \cdot M_x'']\\
    &= \max_{\omega} \max_x  \tr[\mathcal{E}_x [\omega]]
\end{align}
where in the second line we defined a new measurement $M_a' = \tr_{\text{2}}[M_a(\mathbb{1} \ot \omega')]$ and in the third line we defined $M_x'' = \sum_a p(x|a) M_a'$. In the last line we used the fact that  $\max_{M_x} \sum_x \tr[\widetilde{\omega}_x M_x] = \max_x \tr[\widetilde{\omega}_x]$ with $\sum_x M_x = \mathbb{1}$, i.e. we re-expressed the optimization over POVM's with the optimization over $x$.

\section{Complete set of monotones for teleportation simulation}
In this Appendix we show that $q(\mathcal{G}, \Lam{})$ which we defined in (\ref{eq:11}), provides a complete set of monotones for quantum simulation, i.e. all local pre- and post-processings of the the teleportation instrument $\Lam{}$. Similarly, the average success probability $p_{\text{succ}}(\mathbb{E}, \Lam{})$ which we defined in (\ref{eq:100}) provides a complete set of monotones for classical simulation.

 Let us start by focusing on $q(\mathcal{G}, \Lam{})$ and assuming that $\Lam{}$ can be used to simulate $\Lam{}^*$, that is $\Lam{} \geq \Lam{}^*$. We have:
\begin{align}
    \label{eq:109}
    q(\mathcal{G}, \Lam{}) &= \max_{\mathbb{\Lambda}' \leq \mathbb{\Lambda} } \, \max_{\{\mathcal{U}_a\} \in \unitary} \sum_{a} f(a) \tr \left[ (\mathcal{I}\ot \mathcal{U}_a \circ \Lambda_a') \, [\sigma] \cdot \xi_a \right]  \\
    &\geq \max_{\mathbb{\Lambda}' \leq \mathbb{\Lambda}^*} \, \max_{\{\mathcal{U}_a\} \in \unitary} \, \sum_{a} f(a) \tr \left[ (\mathcal{I}\ot \mathcal{U}_a \circ \Lambda_a') \, [\sigma] \cdot \xi_a \right]  \\
    &= q(\mathcal{G}, \Lam{}^*),
\end{align}
since the set $\{\Lam{}' | \Lam{}' \leq \Lam{}^*\}$ is a subset of $\{\Lam{}' | \Lam{}' \leq \Lam{}\}$. We will now assume that $q(\mathcal{G}, \Lam{}) \geq q(\mathcal{G}, \Lam{}^*)$ holds for all games $\mathcal{G} = \{\sigma, \xi_a, f(a)\}$ and show that there always exist a subroutine which allows for the simulation of $\Lam{}^*$ by $\Lam{}$. Let us start by noting that if $q(\mathcal{G}, \Lam{}) \geq q(\mathcal{G}, \Lam{}^*)$ is true for all $\mathcal{G}$ then the following holds:
\begin{align}
    \label{eq:13}
    \forall\,{\mathcal{G}} \qquad \max_{\mathbb{\Lambda}' \leq \mathbb{\Lambda} } \, \max_{\{\mathcal{U}_a\} \in \unitary} \,  \sum_{a} f(a) \tr \left[ (\mathcal{I}\ot \mathcal{U}_a \circ \Lambda_a') \, [\sigma] \cdot \xi_a \right] - \max_{\mathbb{\Lambda}'' \leq \mathbb{\Lambda}^*} \, \max_{\{\mathcal{U}'_b\} \in \unitary} \,  \sum_{b} f(b) \tr \left[ (\mathcal{I}\ot \mathcal{U}_b' \circ \Lambda_b'') \, [\sigma] \cdot \xi_b \right] \geq 0.
\end{align}
Since $\Lam{}'' \leq \Lam{}^*$ we can write $\Lambda_{b}'' = \sum_{a, \lambda} p_{\lambda}\, p(b|a, \lambda)\, \Theta_{\lambda} \circ \Lambda_a^* \circ \Omega_{\lambda}$. Let us now make a particular (and possibly sub-optimal) choice of $p_{\lambda} = \delta_{0, \lambda}$, $p(b|a,\lambda) = \delta_{b, a}$ and $\Theta_{\lambda} = \Omega_{\lambda} = \mathcal{I}$ for all $\lambda$ and let us also choose $\mathcal{U}_b' = \mathcal{U}_b$ for all $b$, i.e. we choose the same correction unitary for both optimization problems. Then (\ref{eq:13}) implies:
\begin{align}
    \label{eq:17}
    \forall\,{\mathcal{G}} \qquad\max_{\mathbb{\Lambda}' \leq \mathbb{\Lambda}} \, \max_{\{\mathcal{U}_a\} \in \unitary}  \sum_a f(a) \tr \left[ (\mathcal{I}\ot \mathcal{U}_a) \left[(\mathcal{I}\ot\Lambda_a' -\mathcal{I}\ot\Lambda_a^* )[\sigma]\right] \cdot \xi_a\right] \geq 0.
\end{align}
We will now claim that (\ref{eq:17}) can only hold if $\Lam{}$ can be used to simulate $\Lam{}^*$. First notice that the above relation must hold for all games $\mathcal{G}$. Hence, let us make a special choice of $\mathcal{G}^* = \{\sigma^*, \xi^*_a, f^*(a)\}$, where:
\begin{align}
    \sigma^* = \frac{\mathbb{1}}{d}  \ot \omega, \quad \xi_a^* = \frac{\mathbb{1}}{d} \ot \mathcal{U}_a^{\dagger}[\eta_a],
\end{align}
and $\mathcal{U}_a^{\dagger}$ is the adjoint of the optimal unitary correction in (\ref{eq:17}). For the moment we assume that $f(a)$, $\omega$ and $\{\eta_a\}$ are arbitrary. Hence, the relation (\ref{eq:17}) implies:
\begin{align}
    \forall\,{f(a), \omega, \{\eta_a\}} \qquad\max_{\mathbb{\Lambda}' \leq \mathbb{\Lambda}} \, \max_{\{\mathcal{U}_a\} \in \unitary}  \sum_a f(a) \tr \left[ (\Lambda_a'[\omega] -\Lambda_a^*[\omega] ) \cdot \eta_a \right] \geq 0.
\end{align}
Denoting $\Delta_a:= \Lambda_a'[\omega] -\Lambda_a^*[\omega]$ we can equivalently write:
\begin{align}   
    \label{eq:delta_mm}
    \forall\,{f(a), \omega, \{\eta_a\}} \qquad\max_{\mathbb{\Lambda}' \leq \mathbb{\Lambda}} \, \max_{\{\mathcal{U}_a\} \in \unitary}  \sum_a f(a) \tr \left[ \Delta_a \cdot \eta_a \right] \geq 0.
\end{align}
Now we will claim that (\ref{eq:delta_mm}) implies that all of the operators $\Delta_a$ are necessary zero. We will prove this by contradiction, i.e. we will start by assuming that there exist a nonzero operator $\Delta_a$ and then show that this leads to a contradiction.

To begin with, notice that due to the no-signalling condition (\ref{eq:14}) the operators $\Delta_a$ satisfy:
\begin{align}
    \sum_a \tr \left[ \Delta_a\right] = 0.
\end{align}
This means that either $(i)$ all operators $\Delta_a$ are identically equal to zero, or $(ii)$ there exist at least one operator $\Delta_{a^*}$ with at least one negative eigenvalue. Suppose for now that $(ii)$ is true and denote this eigenvalue with $\lambda_{a^*} < 0$ and the corresponding eigenvector with $\dyad{\lambda_{a^*}}$. Then in (\ref{eq:delta_mm}) we can choose $f(a^*) = \delta_{a, a^*}$ and $\eta_{a^*} = \dyad{\lambda_{a^*}}$. This would clearly lead to a contradiction with (\ref{eq:delta_mm}).  Hence, we conclude that $(ii)$ cannot be true and the only possibility is that all $\Delta_a = 0$. This equivalently means that $\Lam{}$ can be used to simulate $\Lam{}$, i.e. $\Lam{} \geq \Lam{}^*$.

We now move to the function $p_{\text{succ}}(\mathbb{E}, \Lam{})$ defined in (\ref{eq:100}). Our goal is to show that it constitutes a complete set of monotones for  classical simulation. We proceed analogously as in the case of $q(\mathcal{G}, \Lam{})$. To prove one direction, assume that $p_{\text{succ}}(\mathbb{E}, \Lam{}) \geq p_{\text{succ}}(\mathbb{E}, \Lam{}^*)$ holds for all $\mathbb{E}$. This and the identity (\ref{eq:34}) implies:
\begin{align}
    \label{eq:130}
    \forall\,{\mathbb{E}} \qquad \max_{p(x|a)} \sum_{a,x} p(x|a) \tr\left[(\mathcal{I} \ot \mathcal{E}_x) J_a \cdot \phi_+ \right] - \max_{p'(x|b)} \sum_{b,x} p'(x|b) \tr\left[(\mathcal{I} \ot \mathcal{E}_x) J_b^{\,*} \cdot \phi_+ \right] \geq 0,
\end{align}
where we denoted $J_a = (\mathcal{I}\ot \Lambda_a) \phi_+$ and $J_b^{\,*} = (\mathcal{I}\ot \Lambda_b^*) \phi_+$.
If we now make a particular choice of 
$p'(x|b) = \delta_{x,b}$ for all $b$, then (\ref{eq:130}) implies:
\begin{align}
    \label{eq:131}
    \forall\,{\mathbb{E}} \qquad \max_{p(x|a)} \sum_{x} \tr\left[(\mathcal{I} \ot \mathcal{E}_x) \left(\sum_a p(x|a) J_a - J_x^{\,*} \right) \cdot \phi_+ \right] \geq 0.
\end{align}
We will now claim that (\ref{eq:131}) can only hold if $\Lam{}$ can be used to classically simulate $\Lam{}^*$. To do so, we can define an operator $\Delta_x := \sum_a p(x|a) J_a - J_x^{\,*}  = \sum_a p(x|a) (\mathcal{I}\ot \Lambda_a)\phi_+ - (\mathcal{I}\ot \Lambda_x^*)\phi_+$. Since we have $\sum_x \Delta_x = 0$ we can use analogous arguments as above and infer that (\ref{eq:131}) necessarily implies that $\Delta_x = 0$ for all $x$, or equivalently:
\begin{align}
    \forall\, x \qquad \Lambda_x^* &= \sum_{a} p(x|a) \Lambda_{a}, 
\end{align}
which means that $\Lam{}$ can be used to classicaly simulate $\Lam{}^*$ or equivalently $\Lam{} \succ_c \Lam{}^*$. To prove the reverse direction we assume $\Lam{} \succ_c \Lam{}^*$ which implies that there exist $p(b|a)$ such that $\Lambda_b^* = \sum_a p(b|a) \Lambda_a$ for all $b$. For all games $\mathbb{E}$ we then have:
\begin{align}
    p_{\text{succ}} (\mathbb{E}, \Lam{}^*) &= \max_{p'(x|b)} \sum_{b,x} p'(x|b) \tr \left[ (\mathcal{I} \ot \mathcal{E}_x \circ \Lambda_b^*)\phi_+ \cdot \phi_+ \right] \\
    &= \max_{p'(x|b)} \sum_{a,b,x} p'(x|b) p(b|a) \tr \left[ (\mathcal{I} \ot \mathcal{E}_x \circ \Lambda_a)\phi_+ \cdot \phi_+ \right] \\
    &\leq \max_{p'(x|a)} \sum_{a,x} p'(x|a) \tr \left[ (\mathcal{I} \ot \mathcal{E}_x \circ \Lambda_a)\phi_+ \cdot \phi_+ \right],
\end{align}
where in the last line we defined a new probability distribution $p'(x|a) = \sum_b p'(x|b) p(b|a)$ and inequality follows since this may be not the most general conditional probability distribution.

\end{document}